\documentclass[%
 superscriptaddress,
 reprint,
 showpacs,preprintnumbers,
 nofootinbib,
 amsmath,amssymb,
 aps,
 prl,
 floatfix,
]{revtex4-2}

\usepackage{graphicx}% Include figure files

\usepackage{xcolor}
\usepackage{lineno}
\usepackage{lipsum}
\usepackage{booktabs} % To thicken table lines

\usepackage{amssymb,amsmath,amstext,amsthm,amsfonts,amsfonts}
\usepackage[utf8]{inputenc}
\usepackage{subcaption}
\usepackage[linesnumbered,ruled]{algorithm2e}
\usepackage{url}
\usepackage{soul}
\usepackage{bm}
\usepackage[normalem]{ulem}
\usepackage{multirow, makecell}

\DeclareMathAlphabet{\mathsfit}{\encodingdefault}{\sfdefault}{m}{sl}
\SetMathAlphabet{\mathsfit}{bold}{\encodingdefault}{\sfdefault}{bx}{n}

\newcommand{\appropto}{\mathrel{\vcenter{
  \offinterlineskip\halign{\hfil$##$\cr
    \propto\cr\noalign{\kern2pt}\sim\cr\noalign{\kern-2pt}}}}}

\begin{document}
	\title{Lead perovskites as CE$\nu$NS detectors} 
    \author{C\'esar Jes\'us-Valls}
    \email[E-mail: ]{cesar.jesus-valls@ipmu.jp}
	\affiliation{University of Tokyo, Kavli Institute for the Physics and Mathematics of the Universe (WPI), University of Tokyo Institutes for Advanced Study, Kashiwa, Japan}
    \author{Federico S\'anchez}
	\affiliation{University of Geneva, Section de Physique, DPNC, 1205 Genève, Switzerland}
\begin{abstract}
\noindent
The recent discovery of Coherent Elastic neutrino-Nucleus Scattering (CE$\nu$NS) has created new opportunities to detect and study neutrinos. The interaction cross-section in CE$\nu$NS scales quadratically with the number of neutrons, making heavy-nuclei targets such as active lead-based detectors ideal. In this Letter, we discuss for the first time the potential of semiconductor lead perovskites for building neutrino detectors. Lead perovskites have emerged in the last decade as revolutionary materials for radiation detection due to their heavy and flexible element composition and their unique optoelectronic properties that result in an excellent energy resolution at an economic cost.  While dedicated research and development will be necessary, we find great benefits and no inherent obstacles for the development of lead perovskites as CE$\nu$NS detectors.
\end{abstract}

\maketitle

\section{Introduction}
Neutrinos are the only known fermions carrying exclusively weak charge and therefore are clean probes of the weak interaction and unique messengers of dense matter environments, unaffected by the strong and electromagnetic interactions. This appeals, however, result in a notably suppressed interaction cross-sections, hampering the study of neutrino physics and rendering most applications impractical.\\
In 1974, the existence of coherent elastic neutrino-nucleus scattering (CE$\nu$NS) was pointed out to be a consequence of the Standard Model~\cite{Freedman:1973yd}. In CE$\nu$NS, a neutrino transfers momentum to a whole nucleus via the exchange of a virtual Z boson, forcing it to recoil. The interaction cross-section for this process is
\begin{align}
    \label{eq:cevns}
    \frac{d\sigma^{_{\text{CE}\nu \text{NS}}}}{d E_R} = &\frac{G^2_F}{8\pi\cdot(\hbar c)^4} (N+(1-4\sin^2\theta_W)Z)^2 \nonumber\\
    &\cdot m_N \cdot (2-E_R m_N/E_{\nu}^2)|f(q)|^2\,,
\end{align}
where $G_F$ is the Fermi constant, N (Z) is the number of neutrons (protons),  $\theta_W$ is the Weinberg angle and $m_N$ and $E_R$ are the nucleon mass and its recoil energy respectively.  The nuclear form factor $f(q)$ characterizes the loss of coherence as a function of the transferred momentum $q$=$\sqrt{2m_NE_R}/\hbar$ and it is close to unity for small $q$, associated to typical neutrino energies $E_\nu \lesssim 50$~MeV. Notably, given that $4\sin^2\theta_W\sim 1$ , $\sigma^{_{\text{CE}\nu \text{NS}}}  \appropto N^2$ ~\cite{Scholberg:2020pjn}. This remarkable interaction cross-section enhancement, however, offers a very challenging detection signal, as the nucleon recoil needs to be identified. The maximum recoil energy scales as $E^{\text{max}}_R\approx 2E^2_\nu / m_N$, so that detectors need to be able to measure recoil energies of, at most, several tens of keV. Due to this, it has not been until recently, thanks to progress in detector technology, that CE$\nu$NS have been experimentally demonstrated by the COHERENT collaboration, using
a CsI target~\cite{COHERENT:2017ipa} and an Ar target~\cite{COHERENT:2020iec}. \\
\section{Motivations}
The discovery of  CE$\nu$NS and its enhanced cross-section has the potential to mitigate the elusiveness of neutrinos and therefore to revolutionize its study at energies on the order of a few tens of MeV, which includes geonetrinos~\cite{Sramek:2012nk}, reactor neutrinos~\cite{Qian:2018wid}, accelerator neutrinos from meson decays at rest~\cite{Ajimura:2017fld,Garoby:2017vew, Wang:2013aka, Alonso:2010fs}, solar neutrinos~\cite{Bahcall:2000nu} and supernova neutrino bursts~\cite{Burrows:2020qrp}. Characterizing the cross-section of CE$\nu$NS is also essential for dark matter searches, as CE$\nu$NS constitute an irreducible background, the so-called neutrino floor~\cite{Billard:2013qya}. Being mediated by flavor insensitive neutral currents, the detection of  CE$\nu$NS provides extended sensitivity to sterile neutrinos~\cite{Anderson:2012pn,Dutta:2015nlo,Kosmas:2017zbh} and other new physics signatures~\cite{Krauss:1991ba,Barranco:2007tz,deNiverville:2015mwa,Dutta:2015vwa} and  allows the study of the neutrino magnetic moment~\cite{Dodd:1991ni,Kosmas:2015sqa},  its effective charge radius~\cite{Papavassiliou:2005cs} and the nuclear neutron form factor~\cite{Patton:2012jr,Amanik:2009zz}. Applications, such as deploying neutrino detectors to increase nuclear security~\cite{Stewart:2019rtd,Bernstein:2019hix} might also be possible. Moreover, CE$\nu$NS is relevant to theoretical astrophysics, as a key actor during stellar collapse~\cite{Wilson:1974zz,Schramm:1975nj,Freedman:1977xn}.\\

\section{CE$\nu$NS experiments}
\label{sec:CEVNSexperiments}
Because of all of the above, an increasing number of CE$\nu$NS detector technologies have been proposed~\cite{Drukier:1984vhf,Cabrera:1984rr,Formaggio:2011jt,Braggio:2006dr,Barbeau:2007qi,FernandezMoroni:2014qlq,Horowitz:2003cz,Bondar:2006ma, Joshi:2014fna,RED:2012hpm,Brice:2013fwa,Collar:2014lya} and several experiments are ongoing or have been proposed: COHERENT~\cite{COHERENT:2015mry}, using CsI, NaI, high-purity Ge (HPGe) and liquid-Ar targets; CONUS~\cite{CONNIE:2016ggr}, NCC-1701\cite{Colaresi:2021kus,Colaresi:2022obx} and $\nu$GEN~\cite{Belov:2015ufh} using cryogenic HPGe; MINER~\cite{MINER:2016igy}, using cryogenic HPGe/Si; NUCLEUS~\cite{Strauss:2017cuu} using cryogenic CaWO$_4$ and Al$_2$O$_3$; CONNIE~\cite{CONNIE:2016ggr}, using Si charge coupled devices (CCDs); TEXONO~\cite{Singh:2017jow} using p-type point-contact Ge; RES-NOVA using cryogenic PbWO$_4$~\cite{Pattavina:2020cqc,FerreiroIachellini:2021qgu}; RICOCHET~\cite{Ricochet:2021rjo} using cryogenic HPGe bolometers; and RED100~\cite{RED-100:2019rpf} using liquid-Xe.\\
%This first generation of CE$\nu$NS experiments, however, might be difficult to scale up, hindering the range of future applications. 
To get the most from CE$\nu$NS, an ideal detector should be inexpensive to produce and operate, have excellent energy resolutions to measure nuclear recoils with an energy of a few keV and be made of heavy nuclear target to exploit the quadratic scaling of the cross-section. In this Letter, we point out for the first time the excellent prospects of lead perovskites to build up future CE$\nu$NS detectors.

\section{Lead perovskites}
%
% NOTE PRICE < 75$/kG !!! https://doi.org/10.1038/s41467-019-08981-w
% ESSENTIAL:  CsPbBr3 perovskite detectors with 1.4% energy resolution for high-energy γ-rays
%
\begin{figure}[ht!]
    \centering
    \includegraphics[width=0.99\linewidth]{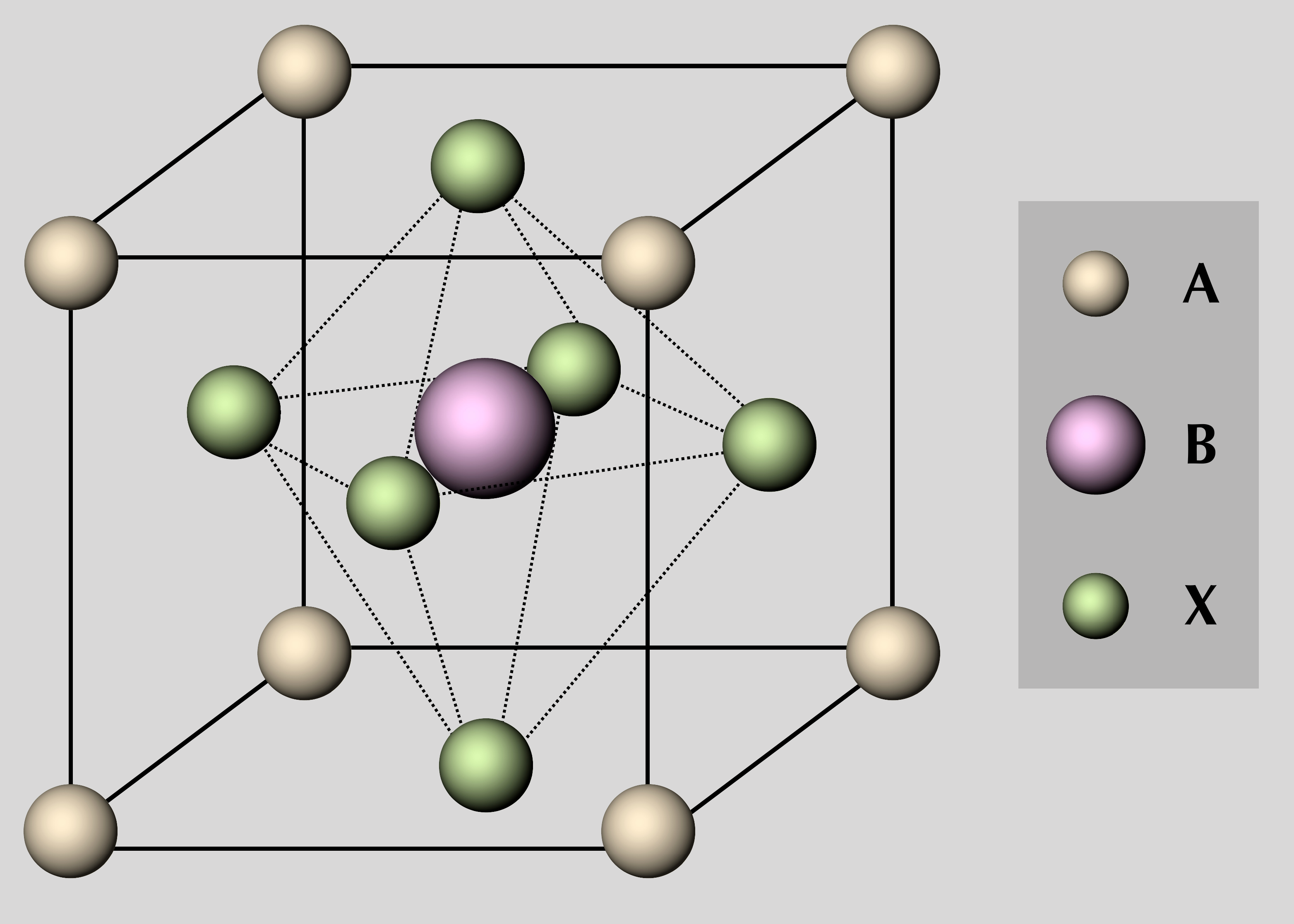}
    \caption{Schematic of the perovskite ABX$_3$ crystal structure.}
    \label{fig:pero_struc}
\end{figure}
Lead halide perovskites (LHP) are novel semiconductors with exceptional optoelectronic properties, versatile chemical composition and low cost synthesis. They typically consist of crystals with structure APbX$_3$, see Fig.~\ref{fig:pero_struc}, where A is CH$_3$NH$_{3+}$ (MA$^+$), CHNH$_{3}^+$ (FA$^+$), or Cs$^+$; B is Pb$^{2+}$; and X is Cl$^-$, Br$^-$, and I$^-$~\cite{wei2019halide}.\\
The study of halide perovskites as photosensors was sparked about a decade ago in the context of solar cell development~\cite{ kojima2009organometal} and quickly emerged as an active field of research due to record energy conversion efficiencies~\cite{green2014emergence,jung2015perovskite,park2015perovskite, correa2017promises, huang2017understanding, yang2018high, kim2020high, yoo2021efficient}. Along the process, much has been learned about the basic properties of this material, which combines a low exciton binding energy on the order of few meV~\cite{miyata2015direct} with exceptionally long electron-hole diffusion lengths exceeding 1 micrometer~\cite{stranks2013electron}, a tunable band gap in the range of 1.2-2.4~eV~\cite{ju2018tunable,unger2017roadmap}, and a high bulk resistivity of 10$^{7-10}\Omega\cdot$cm at room temperature~\cite{pisoni2014ultra}. The combination above is unique, as it pairs efficient charge carrier production and mobility at low voltage bias with a high bulk resistivity, orders of magnitude higher than those of Si and Ge, suppressing dark current and noise. Moreover, LHP naturally allow to manufacture crystals with very high atomic numbers, such as CsPbI$_3$, and to design application-specific perovskite sensors by means of stoichiometry engineering~\cite{emara2016impact,xiao2021grain}. Furthermore, the synthesis of LHP is easy and flexible through techniques such as solution processing. The production cost is also low, with an estimated price of $<0.3$\$/cm$^3$~\cite{wei2019halide}, namely, at the density of 4~g/cm$^3$ an inexpensive cost of $75$\$/kg. Finally, LHP can be operated inexpensively at room temperature.

\section{Existing metrics as $x$/$\gamma$-ray detectors}
Their striking performance as solar cells and their high atomic number\footnote{Photon attenuation increases $\propto Z^4$, where Z is the atomic number.} quickly attracted the interest of the medical imaging community towards lead perovskites~\cite{shrestha2017high, wei2017monolithic, kim2017printable, garcia2017solution, gill2018flexible, zhuang2019highly, zhou2020halide, li2020all, su2020perovskite}. In 2015, MAPbI$_3$ was proven to detect $\gamma$-rays from $^{137}$Cs~\cite{dong2015electron} and first x-ray images were obtained~\cite{yakunin2015detection}. That same year the first detection of single photons in LHP was achieved~\cite{yakunin2015detection}, resolving at room temperature the $E_{\gamma}=59.6$~keV emission from $^{241}$Am with about 35\% resolution at full-width-half-maximum (FWHM). In 2016, the reported X-ray sensitivities at a small bias of 0.1~V were already four-fold those of commercial $\alpha$-Se detectors~\cite{wei2016sensitive}, the dominating material for x-ray imaging. In 2017 perovskites where successfully integrated with CMOS read-out circuitry leading to an almost 30-fold improvement on the x-ray sensitivity~\cite{wei2017monolithic}. In 2017, the spectrum of $^{137}$Cs $\gamma$-ray energy spectrum was obtained, reaching a 6.5\% energy resolution (ER) for the 662~keV peak~\cite{wei2017dopant}. This same metric increased to 3.9\% in 2018~\cite{he2018high} using small 10~mm$^2$ crystals. In 2021, the ER at room temperature for 662~keV photons to 1.4\% while using large CsPbBr$_3$ crystals of 1.5 inches (3.75~cm) in diameter and achieving a stable operation of the sensors for over 18~months~\cite{he2021cspbbr3}. In that study, the spectrum of multiple radioactive sources was resolved, such as 22~keV $^{109}$Cd x-rays at a FWHM ER of 19.7\%. In 2021, promising results were also achieved with CsPbI$_3$, i.e. 20\% ER for 122~keV photons from $^{59}$Co. In 2022, LHP crystals of unprecedented quality have been reported, leading to the best x-ray sensitivities yet achieved in any material~\cite{jiang2022synergistic, he2022sensitivity}.

\section{Prospects as CE$\nu$NS detectors}
\begin{figure}[ht!]
    \centering
    \includegraphics[width=0.99\linewidth]{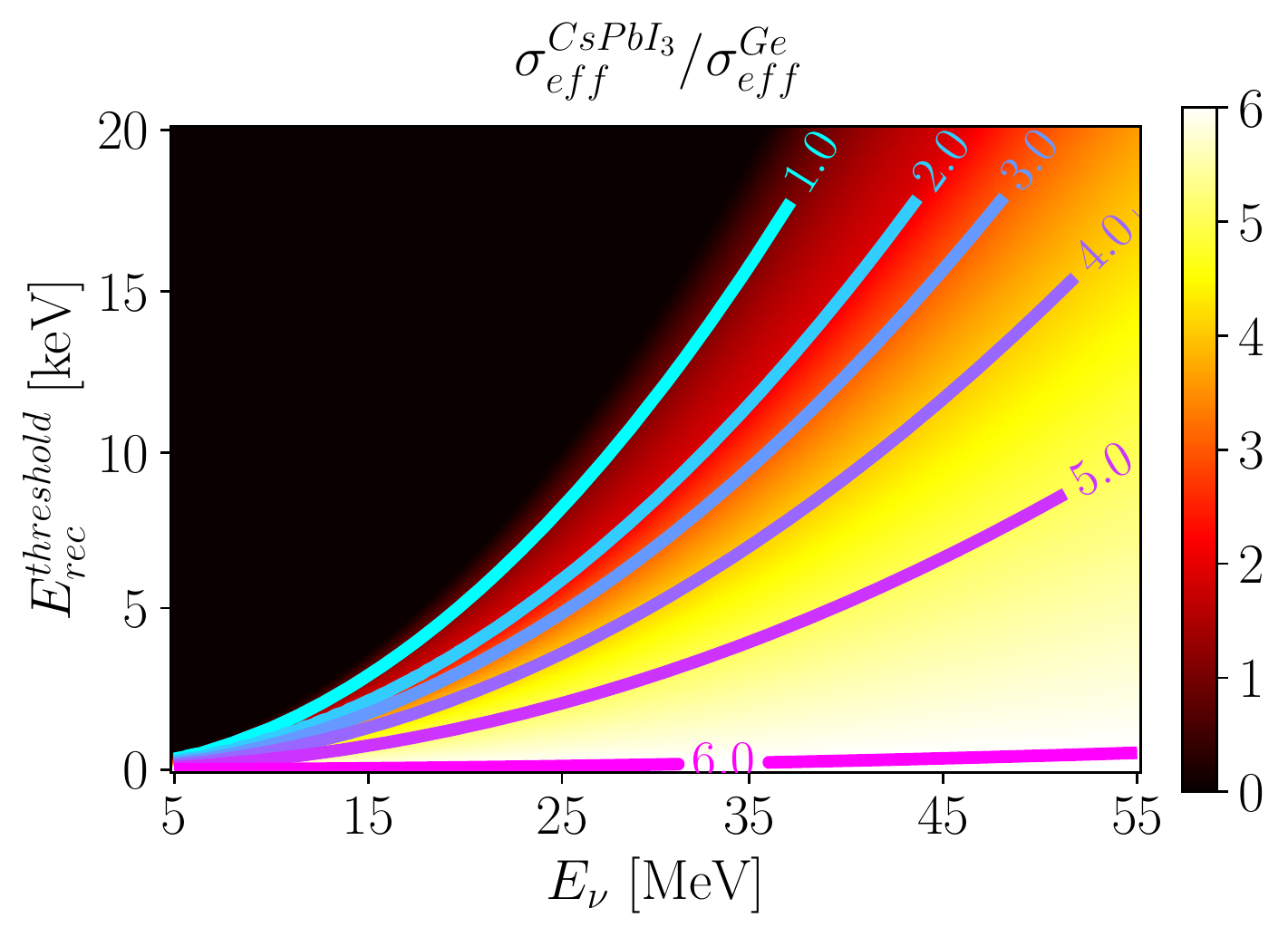}
    \caption{The color map depicts the ratio of $\sigma_{\textup{eff}}^{CsPbI_3}/\sigma_{\textup{eff}}^{Ge}$ as a function of the neutrino energy, $E_\nu$, and the recoil energy threshold, $E^{\textup{recoil}}_{\textup{threshold}}$. A ratio value of zero indicates that $E^{\textup{recoil}}_{\textup{threshold}}$ is above the necessary level to observe any recoil in CsPbI$_3$. To help the visualization, values of some particular integer ratios are highlighted by color lines.}
    \label{fig:sigma_ratio}
\end{figure}

Existing measurement with $x$-rays and $\gamma$-rays are the most abundant tests of lead perovskites as prospective materials to build up radiation detectors. In addition, perovskites-based devices have been already demonstrated to be able to detect $\alpha$~\cite{xie2020lithium} and $\beta$~\cite{yu2020two} particles, as well as neutrons using a hybrid configuration~\cite{andrivcevic2021hybrid}. For the detection of CE$\nu$NS it is important to note that nuclear and electronic recoils have different ionization efficiencies. For Ge it has been measured that nuclear recoils generate about a third of the ionization signal of their electronic counterparts~\cite{EDELWEISS:2003omv}. For lead perovskites this fraction, the so-called quenching, is still unknown. Nonetheless the demonstrated ability of perovskites to resolve energy deposits of few tens of keV is encouraging. The observation of low energy spectral lines, such as $^{59}$Co 22~keV $x$-rays with 19.7\% FWHM ER, suggests that detecting nuclear recoils with energies below 100~keV might be already possible if quenching values are not drastically larger than those in Ge. If lead perovskites are proven to be succesful as nuclear recoil detectors, they have enormous potential as future CE$\nu$NS detectors.\\
Firstly, as earlier reviewed a large fraction of existing or planned CE$\nu$NS detectors use semiconductor materials, typically HPGe, a choice driven by its gold-standard energy resolution. Lead perovskites are rapidly approaching their ultimate energy resolution expected to be similar to that of HPGe~\cite{he2021cspbbr3}. Consequently, in the medium-term LHP might be directly competing with HPGe detectors.
\begin{figure}[ht!]
    \centering
    \includegraphics[width=0.99\linewidth]{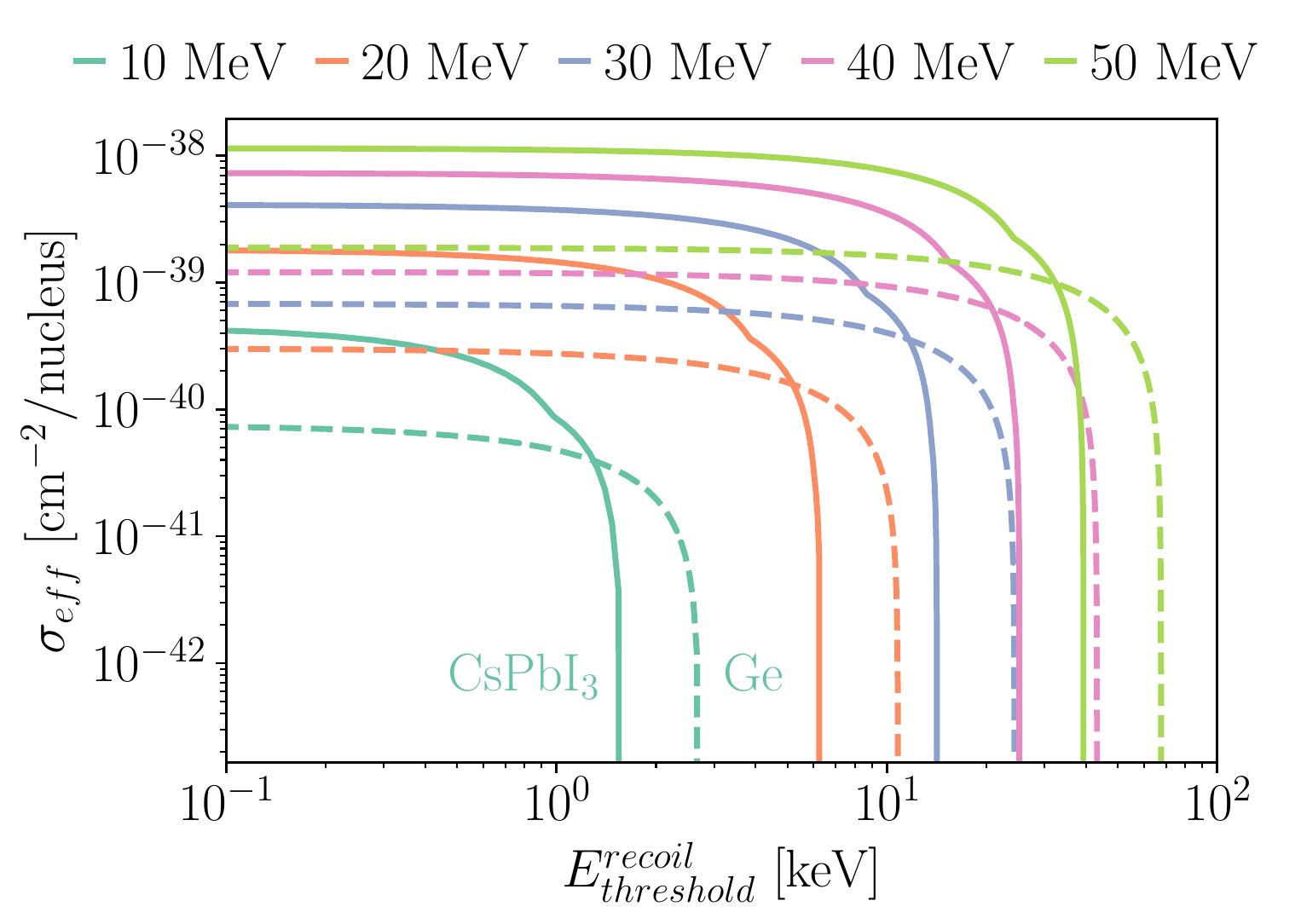}
    \caption{CE$\nu$NS interaction cross-section per nucleus, $\sigma$, multiplied by the detector efficiency, $\epsilon$, as a function of the recoil energy threshold, $E^{\textup{recoil}}_{\textup{threshold}}$. Solid (dashed) lines correspond to CsPbI$_3$ (Ge).} 
    \label{fig:sigma_vs_thresh}
\end{figure}
\\Secondly, producing low activity lead perovskites should be possible, e.g. CsPbI$_3$ consists of Cs and I both used in the first historical detection of CE$\nu$NS~\cite{COHERENT:2017ipa} and archaeological Pb has been recently demonstrated to be adequate for CE$\nu$NS detection~\cite{Beeman:2022wun}. Moreover, CsPbI$_3$ and other lead perovskites are made up of strikingly heavy elements, significantly advantaging the CE$\nu$NS interaction cross-section of mainstream alternative materials and in particular that of Ge. However, the maximum recoil energy decreases linearly with $m_N$ and therefore the ability of the detector to identify the recoiling nucleus needs to be considered. To account for it, we define the effective cross-section, $\sigma_{\textup{eff}}$, as a figure-of-merit, defined as
\begin{align}
\label{eq:sig_eff}
\sigma_{\textup{eff}} \equiv \int_{E^{\textup{recoil}}_{\textup{threshold}}}^{E^{\textup{max}}_{R}} \frac{d\sigma}{dE_R}\,\epsilon\, dE_R \,,
\end{align}
which can be calculated from Eq.~\ref{eq:cevns} if the detector efficiency, $\epsilon$, is specified. Using it, in Fig.\ref{fig:sigma_vs_thresh} CsPbI$_3$ and Ge targets\footnote{For CsPbI$_3$, the weighted average (Cs+Pb+3I)/5 is used on the result of Eq.~\ref{eq:sig_eff}.} are directly compared for some neutrino energies assuming a detector with perfect (null) efficiency above (below) a certain energy recoil threshold, $E^{\textup{recoil}}_{\textup{threshold}}$. As expected, observing interactions in CsPbI$_3$ would require a smaller $E^{\textup{recoil}}_{\textup{threshold}}$ than in Ge, but if the detection threshold is achieved and mildly lowered it results in a large enhancement of the interaction cross-section. This trade-off is characterized by the ratio $\sigma_{\textup{eff}}^{CsPbI_3}/\sigma_{\textup{eff}}^{Ge}$ presented in Fig.\ref{fig:sigma_ratio} as a function of the neutrino energy and the recoil energy threshold.
\\Achieving $E^{\textup{recoil}}_{\textup{threshold}}$ below 20~keV would in general result in an event rate advantage for CsPbI$_3$ detectors compared to Ge when dealing with neutrino energies similar to a few tens of MeV, e.g. neutrinos from pion decays at rest, useful to study sterile neutrinos and NSI, or neutrinos from the high-energy tail of supernova neutrino bursts. Reaching even lower thresholds might result in additional applications involving neutrinos from other sources but it will also likely require a longer term R\&D.\\
Lastly, perovskites are orders of magnitude cheaper to manufacture and potentially to operate\footnote{Perovskites might be able to operate without the need of cryogenic systems, as supported by existing $x$-ray data.} than existing alternatives, including HPGe. Due to their inexpensive production cost, the budget to build a large perovskite detectors would be driven by the number of electronic channels. If volumes similar to a fraction of 1~cm$^3$ could be adequately readout by a single channel, i.e. sizes slightly above of those in use for x-ray detection, a ton of perovskite might be read by $O(10^6)$ electronic channels, making it potentially scalable.
 
\section{Discussion and outlook}
In just one decade lead perovskites have been established as novel materials with transformative potential as radiation detectors due to their unique optoelectronic properties.  In this Letter, we have pointed out their potential as neutrino detectors for the first time and discussed its suitability for the study of CE$\nu$NS. To bring perovskites to their ultimate detection potential and enable their full range of applications active R\&D is required. Demonstrating and characterizing their ability to detect nuclear recoils in the energy range of interest is an essential milestone.\\
Lastly, we note that CE$\nu$NS and some dark-matter models share the same signal mechanism, i.e. the detection of nuclear recoils. Therefore, any progress on that direction might benefit both the neutrino and the dark-matter research communities.

\section*{Acknoledgments}
We acknowledge fruitful discussions with E.~Palomares, and valuable feedback from J.I.~Collar and L.~Pattavina. This project was partially inspired by the ZPro project funded by the Barcelona Institute of Technology (BIST).

\bibliographystyle{apsrev4-1}
\bibliography{biblio}

\begin{thebibliography}{93}%
\makeatletter
\providecommand \@ifxundefined [1]{%
 \@ifx{#1\undefined}
}%
\providecommand \@ifnum [1]{%
 \ifnum #1\expandafter \@firstoftwo
 \else \expandafter \@secondoftwo
 \fi
}%
\providecommand \@ifx [1]{%
 \ifx #1\expandafter \@firstoftwo
 \else \expandafter \@secondoftwo
 \fi
}%
\providecommand \natexlab [1]{#1}%
\providecommand \enquote  [1]{``#1''}%
\providecommand \bibnamefont  [1]{#1}%
\providecommand \bibfnamefont [1]{#1}%
\providecommand \citenamefont [1]{#1}%
\providecommand \href@noop [0]{\@secondoftwo}%
\providecommand \href [0]{\begingroup \@sanitize@url \@href}%
\providecommand \@href[1]{\@@startlink{#1}\@@href}%
\providecommand \@@href[1]{\endgroup#1\@@endlink}%
\providecommand \@sanitize@url [0]{\catcode `\\12\catcode `\$12\catcode
  `\&12\catcode `\#12\catcode `\^12\catcode `\_12\catcode `\%12\relax}%
\providecommand \@@startlink[1]{}%
\providecommand \@@endlink[0]{}%
\providecommand \url  [0]{\begingroup\@sanitize@url \@url }%
\providecommand \@url [1]{\endgroup\@href {#1}{\urlprefix }}%
\providecommand \urlprefix  [0]{URL }%
\providecommand \Eprint [0]{\href }%
\providecommand \doibase [0]{http://dx.doi.org/}%
\providecommand \selectlanguage [0]{\@gobble}%
\providecommand \bibinfo  [0]{\@secondoftwo}%
\providecommand \bibfield  [0]{\@secondoftwo}%
\providecommand \translation [1]{[#1]}%
\providecommand \BibitemOpen [0]{}%
\providecommand \bibitemStop [0]{}%
\providecommand \bibitemNoStop [0]{.\EOS\space}%
\providecommand \EOS [0]{\spacefactor3000\relax}%
\providecommand \BibitemShut  [1]{\csname bibitem#1\endcsname}%
\let\auto@bib@innerbib\@empty
%</preamble>
\bibitem [{\citenamefont {Freedman}(1974)}]{Freedman:1973yd}%
  \BibitemOpen
  \bibfield  {author} {\bibinfo {author} {\bibfnamefont {D.~Z.}\ \bibnamefont
  {Freedman}},\ }\href {\doibase 10.1103/PhysRevD.9.1389} {\bibfield  {journal}
  {\bibinfo  {journal} {Phys. Rev. D}\ }\textbf {\bibinfo {volume} {9}},\
  \bibinfo {pages} {1389} (\bibinfo {year} {1974})}\BibitemShut {NoStop}%
\bibitem [{\citenamefont {Scholberg}(2020)}]{Scholberg:2020pjn}%
  \BibitemOpen
  \bibfield  {author} {\bibinfo {author} {\bibfnamefont {K.}~\bibnamefont
  {Scholberg}},\ }\href {\doibase 10.1088/1742-6596/1468/1/012126} {\bibfield
  {journal} {\bibinfo  {journal} {J. Phys. Conf. Ser.}\ }\textbf {\bibinfo
  {volume} {1468}},\ \bibinfo {pages} {012126} (\bibinfo {year}
  {2020})}\BibitemShut {NoStop}%
\bibitem [{\citenamefont {Akimov}\ \emph {et~al.}(2017)\citenamefont {Akimov}
  \emph {et~al.}}]{COHERENT:2017ipa}%
  \BibitemOpen
  \bibfield  {author} {\bibinfo {author} {\bibfnamefont {D.}~\bibnamefont
  {Akimov}} \emph {et~al.} (\bibinfo {collaboration} {COHERENT}),\ }\href
  {\doibase 10.1126/science.aao0990} {\bibfield  {journal} {\bibinfo  {journal}
  {Science}\ }\textbf {\bibinfo {volume} {357}},\ \bibinfo {pages} {1123}
  (\bibinfo {year} {2017})},\ \Eprint {http://arxiv.org/abs/1708.01294}
  {arXiv:1708.01294 [nucl-ex]} \BibitemShut {NoStop}%
\bibitem [{\citenamefont {Akimov}\ \emph {et~al.}(2021)\citenamefont {Akimov}
  \emph {et~al.}}]{COHERENT:2020iec}%
  \BibitemOpen
  \bibfield  {author} {\bibinfo {author} {\bibfnamefont {D.}~\bibnamefont
  {Akimov}} \emph {et~al.} (\bibinfo {collaboration} {COHERENT}),\ }\href
  {\doibase 10.1103/PhysRevLett.126.012002} {\bibfield  {journal} {\bibinfo
  {journal} {Phys. Rev. Lett.}\ }\textbf {\bibinfo {volume} {126}},\ \bibinfo
  {pages} {012002} (\bibinfo {year} {2021})},\ \Eprint
  {http://arxiv.org/abs/2003.10630} {arXiv:2003.10630 [nucl-ex]} \BibitemShut
  {NoStop}%
\bibitem [{\citenamefont {Sramek}\ \emph {et~al.}(2013)\citenamefont {Sramek},
  \citenamefont {McDonough}, \citenamefont {Kite}, \citenamefont {Lekic},
  \citenamefont {Dye},\ and\ \citenamefont {Zhong}}]{Sramek:2012nk}%
  \BibitemOpen
  \bibfield  {author} {\bibinfo {author} {\bibfnamefont {O.}~\bibnamefont
  {Sramek}}, \bibinfo {author} {\bibfnamefont {W.~F.}\ \bibnamefont
  {McDonough}}, \bibinfo {author} {\bibfnamefont {E.~S.}\ \bibnamefont {Kite}},
  \bibinfo {author} {\bibfnamefont {V.}~\bibnamefont {Lekic}}, \bibinfo
  {author} {\bibfnamefont {S.}~\bibnamefont {Dye}}, \ and\ \bibinfo {author}
  {\bibfnamefont {S.}~\bibnamefont {Zhong}},\ }\href {\doibase
  10.1016/j.epsl.2012.11.001} {\bibfield  {journal} {\bibinfo  {journal} {Earth
  Planet. Sci. Lett.}\ }\textbf {\bibinfo {volume} {361}},\ \bibinfo {pages}
  {356} (\bibinfo {year} {2013})},\ \Eprint {http://arxiv.org/abs/1207.0853}
  {arXiv:1207.0853 [physics.geo-ph]} \BibitemShut {NoStop}%
\bibitem [{\citenamefont {Qian}\ and\ \citenamefont
  {Peng}(2019)}]{Qian:2018wid}%
  \BibitemOpen
  \bibfield  {author} {\bibinfo {author} {\bibfnamefont {X.}~\bibnamefont
  {Qian}}\ and\ \bibinfo {author} {\bibfnamefont {J.-C.}\ \bibnamefont
  {Peng}},\ }\href {\doibase 10.1088/1361-6633/aae881} {\bibfield  {journal}
  {\bibinfo  {journal} {Rept. Prog. Phys.}\ }\textbf {\bibinfo {volume} {82}},\
  \bibinfo {pages} {036201} (\bibinfo {year} {2019})},\ \Eprint
  {http://arxiv.org/abs/1801.05386} {arXiv:1801.05386 [hep-ex]} \BibitemShut
  {NoStop}%
\bibitem [{\citenamefont {Ajimura}\ \emph {et~al.}(2017)\citenamefont {Ajimura}
  \emph {et~al.}}]{Ajimura:2017fld}%
  \BibitemOpen
  \bibfield  {author} {\bibinfo {author} {\bibfnamefont {S.}~\bibnamefont
  {Ajimura}} \emph {et~al.},\ }\href@noop {} {\  (\bibinfo {year} {2017})},\
  \Eprint {http://arxiv.org/abs/1705.08629} {arXiv:1705.08629
  [physics.ins-det]} \BibitemShut {NoStop}%
\bibitem [{\citenamefont {Garoby}\ \emph {et~al.}(2018)\citenamefont {Garoby}
  \emph {et~al.}}]{Garoby:2017vew}%
  \BibitemOpen
  \bibfield  {author} {\bibinfo {author} {\bibfnamefont {R.}~\bibnamefont
  {Garoby}} \emph {et~al.},\ }\href {\doibase 10.1088/1402-4896/aa9bff}
  {\bibfield  {journal} {\bibinfo  {journal} {Phys. Scripta}\ }\textbf
  {\bibinfo {volume} {93}},\ \bibinfo {pages} {014001} (\bibinfo {year}
  {2018})}\BibitemShut {NoStop}%
\bibitem [{\citenamefont {Wang}\ \emph {et~al.}(2013)\citenamefont {Wang},
  \citenamefont {Liang}, \citenamefont {Yin}, \citenamefont {Yu}, \citenamefont
  {He}, \citenamefont {Tao}, \citenamefont {Zhu}, \citenamefont {Jia},\ and\
  \citenamefont {Zhang}}]{Wang:2013aka}%
  \BibitemOpen
  \bibfield  {author} {\bibinfo {author} {\bibfnamefont {F.}~\bibnamefont
  {Wang}}, \bibinfo {author} {\bibfnamefont {T.}~\bibnamefont {Liang}},
  \bibinfo {author} {\bibfnamefont {W.}~\bibnamefont {Yin}}, \bibinfo {author}
  {\bibfnamefont {Q.}~\bibnamefont {Yu}}, \bibinfo {author} {\bibfnamefont
  {L.}~\bibnamefont {He}}, \bibinfo {author} {\bibfnamefont {J.}~\bibnamefont
  {Tao}}, \bibinfo {author} {\bibfnamefont {T.}~\bibnamefont {Zhu}}, \bibinfo
  {author} {\bibfnamefont {X.}~\bibnamefont {Jia}}, \ and\ \bibinfo {author}
  {\bibfnamefont {S.}~\bibnamefont {Zhang}},\ }\href {\doibase
  10.1007/s11433-013-5345-5} {\bibfield  {journal} {\bibinfo  {journal} {Sci.
  China Phys. Mech. Astron.}\ }\textbf {\bibinfo {volume} {56}},\ \bibinfo
  {pages} {2410} (\bibinfo {year} {2013})}\BibitemShut {NoStop}%
\bibitem [{\citenamefont {Alonso}\ \emph {et~al.}(2010)\citenamefont {Alonso}
  \emph {et~al.}}]{Alonso:2010fs}%
  \BibitemOpen
  \bibfield  {author} {\bibinfo {author} {\bibfnamefont {J.}~\bibnamefont
  {Alonso}} \emph {et~al.},\ }\href@noop {} {\  (\bibinfo {year} {2010})},\
  \Eprint {http://arxiv.org/abs/1006.0260} {arXiv:1006.0260 [physics.ins-det]}
  \BibitemShut {NoStop}%
\bibitem [{\citenamefont {Bahcall}\ \emph {et~al.}(2001)\citenamefont
  {Bahcall}, \citenamefont {Pinsonneault},\ and\ \citenamefont
  {Basu}}]{Bahcall:2000nu}%
  \BibitemOpen
  \bibfield  {author} {\bibinfo {author} {\bibfnamefont {J.~N.}\ \bibnamefont
  {Bahcall}}, \bibinfo {author} {\bibfnamefont {M.~H.}\ \bibnamefont
  {Pinsonneault}}, \ and\ \bibinfo {author} {\bibfnamefont {S.}~\bibnamefont
  {Basu}},\ }\href {\doibase 10.1086/321493} {\bibfield  {journal} {\bibinfo
  {journal} {Astrophys. J.}\ }\textbf {\bibinfo {volume} {555}},\ \bibinfo
  {pages} {990} (\bibinfo {year} {2001})},\ \Eprint
  {http://arxiv.org/abs/astro-ph/0010346} {arXiv:astro-ph/0010346} \BibitemShut
  {NoStop}%
\bibitem [{\citenamefont {Burrows}\ and\ \citenamefont
  {Vartanyan}(2021)}]{Burrows:2020qrp}%
  \BibitemOpen
  \bibfield  {author} {\bibinfo {author} {\bibfnamefont {A.}~\bibnamefont
  {Burrows}}\ and\ \bibinfo {author} {\bibfnamefont {D.}~\bibnamefont
  {Vartanyan}},\ }\href {\doibase 10.1038/s41586-020-03059-w} {\bibfield
  {journal} {\bibinfo  {journal} {Nature}\ }\textbf {\bibinfo {volume} {589}},\
  \bibinfo {pages} {29} (\bibinfo {year} {2021})},\ \Eprint
  {http://arxiv.org/abs/2009.14157} {arXiv:2009.14157 [astro-ph.SR]}
  \BibitemShut {NoStop}%
\bibitem [{\citenamefont {Billard}\ \emph {et~al.}(2014)\citenamefont
  {Billard}, \citenamefont {Strigari},\ and\ \citenamefont
  {Figueroa-Feliciano}}]{Billard:2013qya}%
  \BibitemOpen
  \bibfield  {author} {\bibinfo {author} {\bibfnamefont {J.}~\bibnamefont
  {Billard}}, \bibinfo {author} {\bibfnamefont {L.}~\bibnamefont {Strigari}}, \
  and\ \bibinfo {author} {\bibfnamefont {E.}~\bibnamefont
  {Figueroa-Feliciano}},\ }\href {\doibase 10.1103/PhysRevD.89.023524}
  {\bibfield  {journal} {\bibinfo  {journal} {Phys. Rev. D}\ }\textbf {\bibinfo
  {volume} {89}},\ \bibinfo {pages} {023524} (\bibinfo {year} {2014})},\
  \Eprint {http://arxiv.org/abs/1307.5458} {arXiv:1307.5458 [hep-ph]}
  \BibitemShut {NoStop}%
\bibitem [{\citenamefont {Anderson}\ \emph {et~al.}(2012)\citenamefont
  {Anderson}, \citenamefont {Conrad}, \citenamefont {Figueroa-Feliciano},
  \citenamefont {Ignarra}, \citenamefont {Karagiorgi}, \citenamefont
  {Scholberg}, \citenamefont {Shaevitz},\ and\ \citenamefont
  {Spitz}}]{Anderson:2012pn}%
  \BibitemOpen
  \bibfield  {author} {\bibinfo {author} {\bibfnamefont {A.~J.}\ \bibnamefont
  {Anderson}}, \bibinfo {author} {\bibfnamefont {J.~M.}\ \bibnamefont
  {Conrad}}, \bibinfo {author} {\bibfnamefont {E.}~\bibnamefont
  {Figueroa-Feliciano}}, \bibinfo {author} {\bibfnamefont {C.}~\bibnamefont
  {Ignarra}}, \bibinfo {author} {\bibfnamefont {G.}~\bibnamefont {Karagiorgi}},
  \bibinfo {author} {\bibfnamefont {K.}~\bibnamefont {Scholberg}}, \bibinfo
  {author} {\bibfnamefont {M.~H.}\ \bibnamefont {Shaevitz}}, \ and\ \bibinfo
  {author} {\bibfnamefont {J.}~\bibnamefont {Spitz}},\ }\href {\doibase
  10.1103/PhysRevD.86.013004} {\bibfield  {journal} {\bibinfo  {journal} {Phys.
  Rev. D}\ }\textbf {\bibinfo {volume} {86}},\ \bibinfo {pages} {013004}
  (\bibinfo {year} {2012})},\ \Eprint {http://arxiv.org/abs/1201.3805}
  {arXiv:1201.3805 [hep-ph]} \BibitemShut {NoStop}%
\bibitem [{\citenamefont {Dutta}\ \emph
  {et~al.}(2016{\natexlab{a}})\citenamefont {Dutta}, \citenamefont {Gao},
  \citenamefont {Mahapatra}, \citenamefont {Mirabolfathi}, \citenamefont
  {Strigari},\ and\ \citenamefont {Walker}}]{Dutta:2015nlo}%
  \BibitemOpen
  \bibfield  {author} {\bibinfo {author} {\bibfnamefont {B.}~\bibnamefont
  {Dutta}}, \bibinfo {author} {\bibfnamefont {Y.}~\bibnamefont {Gao}}, \bibinfo
  {author} {\bibfnamefont {R.}~\bibnamefont {Mahapatra}}, \bibinfo {author}
  {\bibfnamefont {N.}~\bibnamefont {Mirabolfathi}}, \bibinfo {author}
  {\bibfnamefont {L.~E.}\ \bibnamefont {Strigari}}, \ and\ \bibinfo {author}
  {\bibfnamefont {J.~W.}\ \bibnamefont {Walker}},\ }\href {\doibase
  10.1103/PhysRevD.94.093002} {\bibfield  {journal} {\bibinfo  {journal} {Phys.
  Rev. D}\ }\textbf {\bibinfo {volume} {94}},\ \bibinfo {pages} {093002}
  (\bibinfo {year} {2016}{\natexlab{a}})},\ \Eprint
  {http://arxiv.org/abs/1511.02834} {arXiv:1511.02834 [hep-ph]} \BibitemShut
  {NoStop}%
\bibitem [{\citenamefont {Kosmas}\ \emph {et~al.}(2017)\citenamefont {Kosmas},
  \citenamefont {Papoulias}, \citenamefont {Tortola},\ and\ \citenamefont
  {Valle}}]{Kosmas:2017zbh}%
  \BibitemOpen
  \bibfield  {author} {\bibinfo {author} {\bibfnamefont {T.~S.}\ \bibnamefont
  {Kosmas}}, \bibinfo {author} {\bibfnamefont {D.~K.}\ \bibnamefont
  {Papoulias}}, \bibinfo {author} {\bibfnamefont {M.}~\bibnamefont {Tortola}},
  \ and\ \bibinfo {author} {\bibfnamefont {J.~W.~F.}\ \bibnamefont {Valle}},\
  }\href {\doibase 10.1103/PhysRevD.96.063013} {\bibfield  {journal} {\bibinfo
  {journal} {Phys. Rev. D}\ }\textbf {\bibinfo {volume} {96}},\ \bibinfo
  {pages} {063013} (\bibinfo {year} {2017})},\ \Eprint
  {http://arxiv.org/abs/1703.00054} {arXiv:1703.00054 [hep-ph]} \BibitemShut
  {NoStop}%
\bibitem [{\citenamefont {Krauss}(1991)}]{Krauss:1991ba}%
  \BibitemOpen
  \bibfield  {author} {\bibinfo {author} {\bibfnamefont {L.~M.}\ \bibnamefont
  {Krauss}},\ }\href {\doibase 10.1016/0370-2693(91)90192-S} {\bibfield
  {journal} {\bibinfo  {journal} {Phys. Lett. B}\ }\textbf {\bibinfo {volume}
  {269}},\ \bibinfo {pages} {407} (\bibinfo {year} {1991})}\BibitemShut
  {NoStop}%
\bibitem [{\citenamefont {Barranco}\ \emph {et~al.}(2007)\citenamefont
  {Barranco}, \citenamefont {Miranda},\ and\ \citenamefont
  {Rashba}}]{Barranco:2007tz}%
  \BibitemOpen
  \bibfield  {author} {\bibinfo {author} {\bibfnamefont {J.}~\bibnamefont
  {Barranco}}, \bibinfo {author} {\bibfnamefont {O.~G.}\ \bibnamefont
  {Miranda}}, \ and\ \bibinfo {author} {\bibfnamefont {T.~I.}\ \bibnamefont
  {Rashba}},\ }\href {\doibase 10.1103/PhysRevD.76.073008} {\bibfield
  {journal} {\bibinfo  {journal} {Phys. Rev. D}\ }\textbf {\bibinfo {volume}
  {76}},\ \bibinfo {pages} {073008} (\bibinfo {year} {2007})},\ \Eprint
  {http://arxiv.org/abs/hep-ph/0702175} {arXiv:hep-ph/0702175} \BibitemShut
  {NoStop}%
\bibitem [{\citenamefont {deNiverville}\ \emph {et~al.}(2015)\citenamefont
  {deNiverville}, \citenamefont {Pospelov},\ and\ \citenamefont
  {Ritz}}]{deNiverville:2015mwa}%
  \BibitemOpen
  \bibfield  {author} {\bibinfo {author} {\bibfnamefont {P.}~\bibnamefont
  {deNiverville}}, \bibinfo {author} {\bibfnamefont {M.}~\bibnamefont
  {Pospelov}}, \ and\ \bibinfo {author} {\bibfnamefont {A.}~\bibnamefont
  {Ritz}},\ }\href {\doibase 10.1103/PhysRevD.92.095005} {\bibfield  {journal}
  {\bibinfo  {journal} {Phys. Rev. D}\ }\textbf {\bibinfo {volume} {92}},\
  \bibinfo {pages} {095005} (\bibinfo {year} {2015})},\ \Eprint
  {http://arxiv.org/abs/1505.07805} {arXiv:1505.07805 [hep-ph]} \BibitemShut
  {NoStop}%
\bibitem [{\citenamefont {Dutta}\ \emph
  {et~al.}(2016{\natexlab{b}})\citenamefont {Dutta}, \citenamefont {Mahapatra},
  \citenamefont {Strigari},\ and\ \citenamefont {Walker}}]{Dutta:2015vwa}%
  \BibitemOpen
  \bibfield  {author} {\bibinfo {author} {\bibfnamefont {B.}~\bibnamefont
  {Dutta}}, \bibinfo {author} {\bibfnamefont {R.}~\bibnamefont {Mahapatra}},
  \bibinfo {author} {\bibfnamefont {L.~E.}\ \bibnamefont {Strigari}}, \ and\
  \bibinfo {author} {\bibfnamefont {J.~W.}\ \bibnamefont {Walker}},\ }\href
  {\doibase 10.1103/PhysRevD.93.013015} {\bibfield  {journal} {\bibinfo
  {journal} {Phys. Rev. D}\ }\textbf {\bibinfo {volume} {93}},\ \bibinfo
  {pages} {013015} (\bibinfo {year} {2016}{\natexlab{b}})},\ \Eprint
  {http://arxiv.org/abs/1508.07981} {arXiv:1508.07981 [hep-ph]} \BibitemShut
  {NoStop}%
\bibitem [{\citenamefont {Dodd}\ \emph {et~al.}(1991)\citenamefont {Dodd},
  \citenamefont {Papageorgiu},\ and\ \citenamefont {Ranfone}}]{Dodd:1991ni}%
  \BibitemOpen
  \bibfield  {author} {\bibinfo {author} {\bibfnamefont {A.~C.}\ \bibnamefont
  {Dodd}}, \bibinfo {author} {\bibfnamefont {E.}~\bibnamefont {Papageorgiu}}, \
  and\ \bibinfo {author} {\bibfnamefont {S.}~\bibnamefont {Ranfone}},\ }\href
  {\doibase 10.1016/0370-2693(91)91064-3} {\bibfield  {journal} {\bibinfo
  {journal} {Phys. Lett. B}\ }\textbf {\bibinfo {volume} {266}},\ \bibinfo
  {pages} {434} (\bibinfo {year} {1991})}\BibitemShut {NoStop}%
\bibitem [{\citenamefont {Kosmas}\ \emph {et~al.}(2015)\citenamefont {Kosmas},
  \citenamefont {Miranda}, \citenamefont {Papoulias}, \citenamefont {Tortola},\
  and\ \citenamefont {Valle}}]{Kosmas:2015sqa}%
  \BibitemOpen
  \bibfield  {author} {\bibinfo {author} {\bibfnamefont {T.~S.}\ \bibnamefont
  {Kosmas}}, \bibinfo {author} {\bibfnamefont {O.~G.}\ \bibnamefont {Miranda}},
  \bibinfo {author} {\bibfnamefont {D.~K.}\ \bibnamefont {Papoulias}}, \bibinfo
  {author} {\bibfnamefont {M.}~\bibnamefont {Tortola}}, \ and\ \bibinfo
  {author} {\bibfnamefont {J.~W.~F.}\ \bibnamefont {Valle}},\ }\href {\doibase
  10.1103/PhysRevD.92.013011} {\bibfield  {journal} {\bibinfo  {journal} {Phys.
  Rev. D}\ }\textbf {\bibinfo {volume} {92}},\ \bibinfo {pages} {013011}
  (\bibinfo {year} {2015})},\ \Eprint {http://arxiv.org/abs/1505.03202}
  {arXiv:1505.03202 [hep-ph]} \BibitemShut {NoStop}%
\bibitem [{\citenamefont {Papavassiliou}\ \emph {et~al.}(2006)\citenamefont
  {Papavassiliou}, \citenamefont {Bernabeu},\ and\ \citenamefont
  {Passera}}]{Papavassiliou:2005cs}%
  \BibitemOpen
  \bibfield  {author} {\bibinfo {author} {\bibfnamefont {J.}~\bibnamefont
  {Papavassiliou}}, \bibinfo {author} {\bibfnamefont {J.}~\bibnamefont
  {Bernabeu}}, \ and\ \bibinfo {author} {\bibfnamefont {M.}~\bibnamefont
  {Passera}},\ }\href {\doibase 10.22323/1.021.0192} {\bibfield  {journal}
  {\bibinfo  {journal} {PoS}\ }\textbf {\bibinfo {volume} {HEP2005}},\ \bibinfo
  {pages} {192} (\bibinfo {year} {2006})},\ \Eprint
  {http://arxiv.org/abs/hep-ph/0512029} {arXiv:hep-ph/0512029} \BibitemShut
  {NoStop}%
\bibitem [{\citenamefont {Patton}\ \emph {et~al.}(2012)\citenamefont {Patton},
  \citenamefont {Engel}, \citenamefont {McLaughlin},\ and\ \citenamefont
  {Schunck}}]{Patton:2012jr}%
  \BibitemOpen
  \bibfield  {author} {\bibinfo {author} {\bibfnamefont {K.}~\bibnamefont
  {Patton}}, \bibinfo {author} {\bibfnamefont {J.}~\bibnamefont {Engel}},
  \bibinfo {author} {\bibfnamefont {G.~C.}\ \bibnamefont {McLaughlin}}, \ and\
  \bibinfo {author} {\bibfnamefont {N.}~\bibnamefont {Schunck}},\ }\href
  {\doibase 10.1103/PhysRevC.86.024612} {\bibfield  {journal} {\bibinfo
  {journal} {Phys. Rev. C}\ }\textbf {\bibinfo {volume} {86}},\ \bibinfo
  {pages} {024612} (\bibinfo {year} {2012})},\ \Eprint
  {http://arxiv.org/abs/1207.0693} {arXiv:1207.0693 [nucl-th]} \BibitemShut
  {NoStop}%
\bibitem [{\citenamefont {Amanik}\ and\ \citenamefont
  {McLaughlin}(2009)}]{Amanik:2009zz}%
  \BibitemOpen
  \bibfield  {author} {\bibinfo {author} {\bibfnamefont {P.~S.}\ \bibnamefont
  {Amanik}}\ and\ \bibinfo {author} {\bibfnamefont {G.~C.}\ \bibnamefont
  {McLaughlin}},\ }\href {\doibase 10.1088/0954-3899/36/1/015105} {\bibfield
  {journal} {\bibinfo  {journal} {J. Phys. G}\ }\textbf {\bibinfo {volume}
  {36}},\ \bibinfo {pages} {015105} (\bibinfo {year} {2009})}\BibitemShut
  {NoStop}%
\bibitem [{\citenamefont {Stewart}\ \emph {et~al.}(2019)\citenamefont
  {Stewart}, \citenamefont {Abou-Jaoude},\ and\ \citenamefont
  {Erickson}}]{Stewart:2019rtd}%
  \BibitemOpen
  \bibfield  {author} {\bibinfo {author} {\bibfnamefont {C.}~\bibnamefont
  {Stewart}}, \bibinfo {author} {\bibfnamefont {A.}~\bibnamefont
  {Abou-Jaoude}}, \ and\ \bibinfo {author} {\bibfnamefont {A.}~\bibnamefont
  {Erickson}},\ }\href {\doibase 10.1038/s41467-019-11434-z} {\bibfield
  {journal} {\bibinfo  {journal} {Nature Commun.}\ }\textbf {\bibinfo {volume}
  {10}},\ \bibinfo {pages} {3527} (\bibinfo {year} {2019})}\BibitemShut
  {NoStop}%
\bibitem [{\citenamefont {Bernstein}\ \emph {et~al.}(2020)\citenamefont
  {Bernstein}, \citenamefont {Bowden}, \citenamefont {Goldblum}, \citenamefont
  {Huber}, \citenamefont {Jovanovic},\ and\ \citenamefont
  {Mattingly}}]{Bernstein:2019hix}%
  \BibitemOpen
  \bibfield  {author} {\bibinfo {author} {\bibfnamefont {A.}~\bibnamefont
  {Bernstein}}, \bibinfo {author} {\bibfnamefont {N.}~\bibnamefont {Bowden}},
  \bibinfo {author} {\bibfnamefont {B.~L.}\ \bibnamefont {Goldblum}}, \bibinfo
  {author} {\bibfnamefont {P.}~\bibnamefont {Huber}}, \bibinfo {author}
  {\bibfnamefont {I.}~\bibnamefont {Jovanovic}}, \ and\ \bibinfo {author}
  {\bibfnamefont {J.}~\bibnamefont {Mattingly}},\ }\href {\doibase
  10.1103/RevModPhys.92.011003} {\bibfield  {journal} {\bibinfo  {journal}
  {Rev. Mod. Phys.}\ }\textbf {\bibinfo {volume} {92}},\ \bibinfo {pages}
  {011003} (\bibinfo {year} {2020})},\ \Eprint
  {http://arxiv.org/abs/1908.07113} {arXiv:1908.07113 [physics.soc-ph]}
  \BibitemShut {NoStop}%
\bibitem [{\citenamefont {Wilson}(1974)}]{Wilson:1974zz}%
  \BibitemOpen
  \bibfield  {author} {\bibinfo {author} {\bibfnamefont {J.~R.}\ \bibnamefont
  {Wilson}},\ }\href {\doibase 10.1103/PhysRevLett.32.849} {\bibfield
  {journal} {\bibinfo  {journal} {Phys. Rev. Lett.}\ }\textbf {\bibinfo
  {volume} {32}},\ \bibinfo {pages} {849} (\bibinfo {year} {1974})}\BibitemShut
  {NoStop}%
\bibitem [{\citenamefont {Schramm}\ and\ \citenamefont
  {Arnett}(1975)}]{Schramm:1975nj}%
  \BibitemOpen
  \bibfield  {author} {\bibinfo {author} {\bibfnamefont {D.~N.}\ \bibnamefont
  {Schramm}}\ and\ \bibinfo {author} {\bibfnamefont {W.~D.}\ \bibnamefont
  {Arnett}},\ }\href {\doibase 10.1103/PhysRevLett.34.113} {\bibfield
  {journal} {\bibinfo  {journal} {Phys. Rev. Lett.}\ }\textbf {\bibinfo
  {volume} {34}},\ \bibinfo {pages} {113} (\bibinfo {year} {1975})}\BibitemShut
  {NoStop}%
\bibitem [{\citenamefont {Freedman}\ \emph {et~al.}(1977)\citenamefont
  {Freedman}, \citenamefont {Schramm},\ and\ \citenamefont
  {Tubbs}}]{Freedman:1977xn}%
  \BibitemOpen
  \bibfield  {author} {\bibinfo {author} {\bibfnamefont {D.~Z.}\ \bibnamefont
  {Freedman}}, \bibinfo {author} {\bibfnamefont {D.~N.}\ \bibnamefont
  {Schramm}}, \ and\ \bibinfo {author} {\bibfnamefont {D.~L.}\ \bibnamefont
  {Tubbs}},\ }\href {\doibase 10.1146/annurev.ns.27.120177.001123} {\bibfield
  {journal} {\bibinfo  {journal} {Ann. Rev. Nucl. Part. Sci.}\ }\textbf
  {\bibinfo {volume} {27}},\ \bibinfo {pages} {167} (\bibinfo {year}
  {1977})}\BibitemShut {NoStop}%
\bibitem [{\citenamefont {Drukier}\ and\ \citenamefont
  {Stodolsky}(1984)}]{Drukier:1984vhf}%
  \BibitemOpen
  \bibfield  {author} {\bibinfo {author} {\bibfnamefont {A.}~\bibnamefont
  {Drukier}}\ and\ \bibinfo {author} {\bibfnamefont {L.}~\bibnamefont
  {Stodolsky}},\ }\href {\doibase 10.1103/PhysRevD.30.2295} {\bibfield
  {journal} {\bibinfo  {journal} {Phys. Rev. D}\ }\textbf {\bibinfo {volume}
  {30}},\ \bibinfo {pages} {2295} (\bibinfo {year} {1984})}\BibitemShut
  {NoStop}%
\bibitem [{\citenamefont {Cabrera}\ \emph {et~al.}(1985)\citenamefont
  {Cabrera}, \citenamefont {Krauss},\ and\ \citenamefont
  {Wilczek}}]{Cabrera:1984rr}%
  \BibitemOpen
  \bibfield  {author} {\bibinfo {author} {\bibfnamefont {B.}~\bibnamefont
  {Cabrera}}, \bibinfo {author} {\bibfnamefont {L.~M.}\ \bibnamefont {Krauss}},
  \ and\ \bibinfo {author} {\bibfnamefont {F.}~\bibnamefont {Wilczek}},\ }\href
  {\doibase 10.1103/PhysRevLett.55.25} {\bibfield  {journal} {\bibinfo
  {journal} {Phys. Rev. Lett.}\ }\textbf {\bibinfo {volume} {55}},\ \bibinfo
  {pages} {25} (\bibinfo {year} {1985})}\BibitemShut {NoStop}%
\bibitem [{\citenamefont {Formaggio}\ \emph {et~al.}(2012)\citenamefont
  {Formaggio}, \citenamefont {Figueroa-Feliciano},\ and\ \citenamefont
  {Anderson}}]{Formaggio:2011jt}%
  \BibitemOpen
  \bibfield  {author} {\bibinfo {author} {\bibfnamefont {J.~A.}\ \bibnamefont
  {Formaggio}}, \bibinfo {author} {\bibfnamefont {E.}~\bibnamefont
  {Figueroa-Feliciano}}, \ and\ \bibinfo {author} {\bibfnamefont {A.~J.}\
  \bibnamefont {Anderson}},\ }\href {\doibase 10.1103/PhysRevD.85.013009}
  {\bibfield  {journal} {\bibinfo  {journal} {Phys. Rev. D}\ }\textbf {\bibinfo
  {volume} {85}},\ \bibinfo {pages} {013009} (\bibinfo {year} {2012})},\
  \Eprint {http://arxiv.org/abs/1107.3512} {arXiv:1107.3512 [hep-ph]}
  \BibitemShut {NoStop}%
\bibitem [{\citenamefont {Braggio}\ \emph {et~al.}(2006)\citenamefont
  {Braggio}, \citenamefont {Bressi}, \citenamefont {Carugno}, \citenamefont
  {Feltrin},\ and\ \citenamefont {Galeazzi}}]{Braggio:2006dr}%
  \BibitemOpen
  \bibfield  {author} {\bibinfo {author} {\bibfnamefont {C.}~\bibnamefont
  {Braggio}}, \bibinfo {author} {\bibfnamefont {G.}~\bibnamefont {Bressi}},
  \bibinfo {author} {\bibfnamefont {G.}~\bibnamefont {Carugno}}, \bibinfo
  {author} {\bibfnamefont {E.}~\bibnamefont {Feltrin}}, \ and\ \bibinfo
  {author} {\bibfnamefont {G.}~\bibnamefont {Galeazzi}},\ }\href {\doibase
  10.1016/j.nima.2006.06.008} {\bibfield  {journal} {\bibinfo  {journal} {Nucl.
  Instrum. Meth. A}\ }\textbf {\bibinfo {volume} {568}},\ \bibinfo {pages}
  {412} (\bibinfo {year} {2006})}\BibitemShut {NoStop}%
\bibitem [{\citenamefont {Barbeau}\ \emph {et~al.}(2007)\citenamefont
  {Barbeau}, \citenamefont {Collar},\ and\ \citenamefont
  {Tench}}]{Barbeau:2007qi}%
  \BibitemOpen
  \bibfield  {author} {\bibinfo {author} {\bibfnamefont {P.~S.}\ \bibnamefont
  {Barbeau}}, \bibinfo {author} {\bibfnamefont {J.~I.}\ \bibnamefont {Collar}},
  \ and\ \bibinfo {author} {\bibfnamefont {O.}~\bibnamefont {Tench}},\ }\href
  {\doibase 10.1088/1475-7516/2007/09/009} {\bibfield  {journal} {\bibinfo
  {journal} {JCAP}\ }\textbf {\bibinfo {volume} {09}},\ \bibinfo {pages} {009}
  (\bibinfo {year} {2007})},\ \Eprint {http://arxiv.org/abs/nucl-ex/0701012}
  {arXiv:nucl-ex/0701012} \BibitemShut {NoStop}%
\bibitem [{\citenamefont {Fernandez~Moroni}\ \emph {et~al.}(2015)\citenamefont
  {Fernandez~Moroni}, \citenamefont {Estrada}, \citenamefont {Paolini},
  \citenamefont {Cancelo}, \citenamefont {Tiffenberg},\ and\ \citenamefont
  {Molina}}]{FernandezMoroni:2014qlq}%
  \BibitemOpen
  \bibfield  {author} {\bibinfo {author} {\bibfnamefont {G.}~\bibnamefont
  {Fernandez~Moroni}}, \bibinfo {author} {\bibfnamefont {J.}~\bibnamefont
  {Estrada}}, \bibinfo {author} {\bibfnamefont {E.~E.}\ \bibnamefont
  {Paolini}}, \bibinfo {author} {\bibfnamefont {G.}~\bibnamefont {Cancelo}},
  \bibinfo {author} {\bibfnamefont {J.}~\bibnamefont {Tiffenberg}}, \ and\
  \bibinfo {author} {\bibfnamefont {J.}~\bibnamefont {Molina}},\ }\href
  {\doibase 10.1103/PhysRevD.91.072001} {\bibfield  {journal} {\bibinfo
  {journal} {Phys. Rev. D}\ }\textbf {\bibinfo {volume} {91}},\ \bibinfo
  {pages} {072001} (\bibinfo {year} {2015})},\ \Eprint
  {http://arxiv.org/abs/1405.5761} {arXiv:1405.5761 [physics.ins-det]}
  \BibitemShut {NoStop}%
\bibitem [{\citenamefont {Horowitz}\ \emph {et~al.}(2003)\citenamefont
  {Horowitz}, \citenamefont {Coakley},\ and\ \citenamefont
  {McKinsey}}]{Horowitz:2003cz}%
  \BibitemOpen
  \bibfield  {author} {\bibinfo {author} {\bibfnamefont {C.~J.}\ \bibnamefont
  {Horowitz}}, \bibinfo {author} {\bibfnamefont {K.~J.}\ \bibnamefont
  {Coakley}}, \ and\ \bibinfo {author} {\bibfnamefont {D.~N.}\ \bibnamefont
  {McKinsey}},\ }\href {\doibase 10.1103/PhysRevD.68.023005} {\bibfield
  {journal} {\bibinfo  {journal} {Phys. Rev. D}\ }\textbf {\bibinfo {volume}
  {68}},\ \bibinfo {pages} {023005} (\bibinfo {year} {2003})},\ \Eprint
  {http://arxiv.org/abs/astro-ph/0302071} {arXiv:astro-ph/0302071} \BibitemShut
  {NoStop}%
\bibitem [{\citenamefont {Bondar}\ \emph {et~al.}(2007)\citenamefont {Bondar},
  \citenamefont {Buzulutskov}, \citenamefont {Grebenuk}, \citenamefont
  {Pavlyuchenko}, \citenamefont {Snopkov}, \citenamefont {Tikhonov},
  \citenamefont {Kudryavtsev}, \citenamefont {Lightfoot},\ and\ \citenamefont
  {Spooner}}]{Bondar:2006ma}%
  \BibitemOpen
  \bibfield  {author} {\bibinfo {author} {\bibfnamefont {A.}~\bibnamefont
  {Bondar}}, \bibinfo {author} {\bibfnamefont {A.}~\bibnamefont {Buzulutskov}},
  \bibinfo {author} {\bibfnamefont {A.}~\bibnamefont {Grebenuk}}, \bibinfo
  {author} {\bibfnamefont {D.}~\bibnamefont {Pavlyuchenko}}, \bibinfo {author}
  {\bibfnamefont {R.}~\bibnamefont {Snopkov}}, \bibinfo {author} {\bibfnamefont
  {Y.}~\bibnamefont {Tikhonov}}, \bibinfo {author} {\bibfnamefont {V.~A.}\
  \bibnamefont {Kudryavtsev}}, \bibinfo {author} {\bibfnamefont {P.~K.}\
  \bibnamefont {Lightfoot}}, \ and\ \bibinfo {author} {\bibfnamefont
  {N.~J.~C.}\ \bibnamefont {Spooner}},\ }\href {\doibase
  10.1016/j.nima.2007.01.090} {\bibfield  {journal} {\bibinfo  {journal} {Nucl.
  Instrum. Meth. A}\ }\textbf {\bibinfo {volume} {574}},\ \bibinfo {pages}
  {493} (\bibinfo {year} {2007})},\ \Eprint
  {http://arxiv.org/abs/physics/0611068} {arXiv:physics/0611068} \BibitemShut
  {NoStop}%
\bibitem [{\citenamefont {Joshi}\ \emph {et~al.}(2014)\citenamefont {Joshi}
  \emph {et~al.}}]{Joshi:2014fna}%
  \BibitemOpen
  \bibfield  {author} {\bibinfo {author} {\bibfnamefont {T.~H.}\ \bibnamefont
  {Joshi}} \emph {et~al.},\ }\href {\doibase 10.1103/PhysRevLett.112.171303}
  {\bibfield  {journal} {\bibinfo  {journal} {Phys. Rev. Lett.}\ }\textbf
  {\bibinfo {volume} {112}},\ \bibinfo {pages} {171303} (\bibinfo {year}
  {2014})},\ \Eprint {http://arxiv.org/abs/1402.2037} {arXiv:1402.2037
  [physics.ins-det]} \BibitemShut {NoStop}%
\bibitem [{\citenamefont {Akimov}\ \emph {et~al.}(2013)\citenamefont {Akimov}
  \emph {et~al.}}]{RED:2012hpm}%
  \BibitemOpen
  \bibfield  {author} {\bibinfo {author} {\bibfnamefont {D.~Y.}\ \bibnamefont
  {Akimov}} \emph {et~al.} (\bibinfo {collaboration} {RED}),\ }\href {\doibase
  10.1088/1748-0221/8/10/P10023} {\bibfield  {journal} {\bibinfo  {journal}
  {JINST}\ }\textbf {\bibinfo {volume} {8}},\ \bibinfo {pages} {P10023}
  (\bibinfo {year} {2013})},\ \Eprint {http://arxiv.org/abs/1212.1938}
  {arXiv:1212.1938 [physics.ins-det]} \BibitemShut {NoStop}%
\bibitem [{\citenamefont {Brice}\ \emph {et~al.}(2014)\citenamefont {Brice}
  \emph {et~al.}}]{Brice:2013fwa}%
  \BibitemOpen
  \bibfield  {author} {\bibinfo {author} {\bibfnamefont {S.~J.}\ \bibnamefont
  {Brice}} \emph {et~al.},\ }\href {\doibase 10.1103/PhysRevD.89.072004}
  {\bibfield  {journal} {\bibinfo  {journal} {Phys. Rev. D}\ }\textbf {\bibinfo
  {volume} {89}},\ \bibinfo {pages} {072004} (\bibinfo {year} {2014})},\
  \Eprint {http://arxiv.org/abs/1311.5958} {arXiv:1311.5958 [physics.ins-det]}
  \BibitemShut {NoStop}%
\bibitem [{\citenamefont {Collar}\ \emph {et~al.}(2015)\citenamefont {Collar},
  \citenamefont {Fields}, \citenamefont {Hai}, \citenamefont {Hossbach},
  \citenamefont {Orrell}, \citenamefont {Overman}, \citenamefont {Perumpilly},\
  and\ \citenamefont {Scholz}}]{Collar:2014lya}%
  \BibitemOpen
  \bibfield  {author} {\bibinfo {author} {\bibfnamefont {J.~I.}\ \bibnamefont
  {Collar}}, \bibinfo {author} {\bibfnamefont {N.~E.}\ \bibnamefont {Fields}},
  \bibinfo {author} {\bibfnamefont {M.}~\bibnamefont {Hai}}, \bibinfo {author}
  {\bibfnamefont {T.~W.}\ \bibnamefont {Hossbach}}, \bibinfo {author}
  {\bibfnamefont {J.~L.}\ \bibnamefont {Orrell}}, \bibinfo {author}
  {\bibfnamefont {C.~T.}\ \bibnamefont {Overman}}, \bibinfo {author}
  {\bibfnamefont {G.}~\bibnamefont {Perumpilly}}, \ and\ \bibinfo {author}
  {\bibfnamefont {B.}~\bibnamefont {Scholz}},\ }\href {\doibase
  10.1016/j.nima.2014.11.037} {\bibfield  {journal} {\bibinfo  {journal} {Nucl.
  Instrum. Meth. A}\ }\textbf {\bibinfo {volume} {773}},\ \bibinfo {pages} {56}
  (\bibinfo {year} {2015})},\ \Eprint {http://arxiv.org/abs/1407.7524}
  {arXiv:1407.7524 [physics.ins-det]} \BibitemShut {NoStop}%
\bibitem [{\citenamefont {Akimov}\ \emph {et~al.}(2015)\citenamefont {Akimov}
  \emph {et~al.}}]{COHERENT:2015mry}%
  \BibitemOpen
  \bibfield  {author} {\bibinfo {author} {\bibfnamefont {D.}~\bibnamefont
  {Akimov}} \emph {et~al.} (\bibinfo {collaboration} {COHERENT}),\ }\href@noop
  {} {\  (\bibinfo {year} {2015})},\ \Eprint {http://arxiv.org/abs/1509.08702}
  {arXiv:1509.08702 [physics.ins-det]} \BibitemShut {NoStop}%
\bibitem [{\citenamefont {Aguilar-Arevalo}\ \emph {et~al.}(2016)\citenamefont
  {Aguilar-Arevalo} \emph {et~al.}}]{CONNIE:2016ggr}%
  \BibitemOpen
  \bibfield  {author} {\bibinfo {author} {\bibfnamefont {A.}~\bibnamefont
  {Aguilar-Arevalo}} \emph {et~al.} (\bibinfo {collaboration} {CONNIE}),\
  }\href {\doibase 10.1088/1742-6596/761/1/012057} {\bibfield  {journal}
  {\bibinfo  {journal} {J. Phys. Conf. Ser.}\ }\textbf {\bibinfo {volume}
  {761}},\ \bibinfo {pages} {012057} (\bibinfo {year} {2016})},\ \Eprint
  {http://arxiv.org/abs/1608.01565} {arXiv:1608.01565 [physics.ins-det]}
  \BibitemShut {NoStop}%
\bibitem [{\citenamefont {Colaresi}\ \emph {et~al.}(2021)\citenamefont
  {Colaresi}, \citenamefont {Collar}, \citenamefont {Hossbach}, \citenamefont
  {Kavner}, \citenamefont {Lewis}, \citenamefont {Robinson},\ and\
  \citenamefont {Yocum}}]{Colaresi:2021kus}%
  \BibitemOpen
  \bibfield  {author} {\bibinfo {author} {\bibfnamefont {J.}~\bibnamefont
  {Colaresi}}, \bibinfo {author} {\bibfnamefont {J.~I.}\ \bibnamefont
  {Collar}}, \bibinfo {author} {\bibfnamefont {T.~W.}\ \bibnamefont
  {Hossbach}}, \bibinfo {author} {\bibfnamefont {A.~R.~L.}\ \bibnamefont
  {Kavner}}, \bibinfo {author} {\bibfnamefont {C.~M.}\ \bibnamefont {Lewis}},
  \bibinfo {author} {\bibfnamefont {A.~E.}\ \bibnamefont {Robinson}}, \ and\
  \bibinfo {author} {\bibfnamefont {K.~M.}\ \bibnamefont {Yocum}},\ }\href
  {\doibase 10.1103/PhysRevD.104.072003} {\bibfield  {journal} {\bibinfo
  {journal} {Phys. Rev. D}\ }\textbf {\bibinfo {volume} {104}},\ \bibinfo
  {pages} {072003} (\bibinfo {year} {2021})},\ \Eprint
  {http://arxiv.org/abs/2108.02880} {arXiv:2108.02880 [hep-ex]} \BibitemShut
  {NoStop}%
\bibitem [{\citenamefont {Colaresi}\ \emph {et~al.}(2022)\citenamefont
  {Colaresi}, \citenamefont {Collar}, \citenamefont {Hossbach}, \citenamefont
  {Lewis},\ and\ \citenamefont {Yocum}}]{Colaresi:2022obx}%
  \BibitemOpen
  \bibfield  {author} {\bibinfo {author} {\bibfnamefont {J.}~\bibnamefont
  {Colaresi}}, \bibinfo {author} {\bibfnamefont {J.~I.}\ \bibnamefont
  {Collar}}, \bibinfo {author} {\bibfnamefont {T.~W.}\ \bibnamefont
  {Hossbach}}, \bibinfo {author} {\bibfnamefont {C.~M.}\ \bibnamefont {Lewis}},
  \ and\ \bibinfo {author} {\bibfnamefont {K.~M.}\ \bibnamefont {Yocum}},\
  }\href {\doibase 10.1103/PhysRevLett.129.211802} {\bibfield  {journal}
  {\bibinfo  {journal} {Phys. Rev. Lett.}\ }\textbf {\bibinfo {volume} {129}},\
  \bibinfo {pages} {211802} (\bibinfo {year} {2022})},\ \Eprint
  {http://arxiv.org/abs/2202.09672} {arXiv:2202.09672 [hep-ex]} \BibitemShut
  {NoStop}%
\bibitem [{\citenamefont {Belov}\ \emph {et~al.}(2015)\citenamefont {Belov}
  \emph {et~al.}}]{Belov:2015ufh}%
  \BibitemOpen
  \bibfield  {author} {\bibinfo {author} {\bibfnamefont {V.}~\bibnamefont
  {Belov}} \emph {et~al.},\ }\href {\doibase 10.1088/1748-0221/10/12/P12011}
  {\bibfield  {journal} {\bibinfo  {journal} {JINST}\ }\textbf {\bibinfo
  {volume} {10}},\ \bibinfo {pages} {P12011} (\bibinfo {year}
  {2015})}\BibitemShut {NoStop}%
\bibitem [{\citenamefont {Agnolet}\ \emph {et~al.}(2017)\citenamefont {Agnolet}
  \emph {et~al.}}]{MINER:2016igy}%
  \BibitemOpen
  \bibfield  {author} {\bibinfo {author} {\bibfnamefont {G.}~\bibnamefont
  {Agnolet}} \emph {et~al.} (\bibinfo {collaboration} {MINER}),\ }\href
  {\doibase 10.1016/j.nima.2017.02.024} {\bibfield  {journal} {\bibinfo
  {journal} {Nucl. Instrum. Meth. A}\ }\textbf {\bibinfo {volume} {853}},\
  \bibinfo {pages} {53} (\bibinfo {year} {2017})},\ \Eprint
  {http://arxiv.org/abs/1609.02066} {arXiv:1609.02066 [physics.ins-det]}
  \BibitemShut {NoStop}%
\bibitem [{\citenamefont {Strauss}\ \emph {et~al.}(2017)\citenamefont {Strauss}
  \emph {et~al.}}]{Strauss:2017cuu}%
  \BibitemOpen
  \bibfield  {author} {\bibinfo {author} {\bibfnamefont {R.}~\bibnamefont
  {Strauss}} \emph {et~al.},\ }\href {\doibase 10.1140/epjc/s10052-017-5068-2}
  {\bibfield  {journal} {\bibinfo  {journal} {Eur. Phys. J. C}\ }\textbf
  {\bibinfo {volume} {77}},\ \bibinfo {pages} {506} (\bibinfo {year} {2017})},\
  \Eprint {http://arxiv.org/abs/1704.04320} {arXiv:1704.04320
  [physics.ins-det]} \BibitemShut {NoStop}%
\bibitem [{\citenamefont {Singh}\ and\ \citenamefont
  {Wong}(2017)}]{Singh:2017jow}%
  \BibitemOpen
  \bibfield  {author} {\bibinfo {author} {\bibfnamefont {L.}~\bibnamefont
  {Singh}}\ and\ \bibinfo {author} {\bibfnamefont {H.~T.}\ \bibnamefont {Wong}}
  (\bibinfo {collaboration} {TEXONO}),\ }\href {\doibase
  10.1088/1742-6596/888/1/012124} {\bibfield  {journal} {\bibinfo  {journal}
  {J. Phys. Conf. Ser.}\ }\textbf {\bibinfo {volume} {888}},\ \bibinfo {pages}
  {012124} (\bibinfo {year} {2017})}\BibitemShut {NoStop}%
\bibitem [{\citenamefont {Pattavina}\ \emph {et~al.}(2020)\citenamefont
  {Pattavina}, \citenamefont {Ferreiro~Iachellini},\ and\ \citenamefont
  {Tamborra}}]{Pattavina:2020cqc}%
  \BibitemOpen
  \bibfield  {author} {\bibinfo {author} {\bibfnamefont {L.}~\bibnamefont
  {Pattavina}}, \bibinfo {author} {\bibfnamefont {N.}~\bibnamefont
  {Ferreiro~Iachellini}}, \ and\ \bibinfo {author} {\bibfnamefont
  {I.}~\bibnamefont {Tamborra}},\ }\href {\doibase 10.1103/PhysRevD.102.063001}
  {\bibfield  {journal} {\bibinfo  {journal} {Phys. Rev. D}\ }\textbf {\bibinfo
  {volume} {102}},\ \bibinfo {pages} {063001} (\bibinfo {year} {2020})},\
  \Eprint {http://arxiv.org/abs/2004.06936} {arXiv:2004.06936 [astro-ph.HE]}
  \BibitemShut {NoStop}%
\bibitem [{\citenamefont {Ferreiro~Iachellini}\ \emph
  {et~al.}(2022)\citenamefont {Ferreiro~Iachellini} \emph
  {et~al.}}]{FerreiroIachellini:2021qgu}%
  \BibitemOpen
  \bibfield  {author} {\bibinfo {author} {\bibfnamefont {N.}~\bibnamefont
  {Ferreiro~Iachellini}} \emph {et~al.},\ }\href {\doibase
  10.1007/s10909-022-02823-8} {\bibfield  {journal} {\bibinfo  {journal} {J.
  Low Temp. Phys.}\ }\textbf {\bibinfo {volume} {209}},\ \bibinfo {pages} {872}
  (\bibinfo {year} {2022})},\ \Eprint {http://arxiv.org/abs/2111.07638}
  {arXiv:2111.07638 [physics.ins-det]} \BibitemShut {NoStop}%
\bibitem [{\citenamefont {Augier}\ \emph {et~al.}(2021)\citenamefont {Augier}
  \emph {et~al.}}]{Ricochet:2021rjo}%
  \BibitemOpen
  \bibfield  {author} {\bibinfo {author} {\bibfnamefont {C.}~\bibnamefont
  {Augier}} \emph {et~al.} (\bibinfo {collaboration} {Ricochet}),\ }in\
  \href@noop {} {\emph {\bibinfo {booktitle} {{19th International Workshop on
  Low Temperature Detectors}}}}\ (\bibinfo {year} {2021})\ \Eprint
  {http://arxiv.org/abs/2111.06745} {arXiv:2111.06745 [physics.ins-det]}
  \BibitemShut {NoStop}%
\bibitem [{\citenamefont {Akimov}\ \emph {et~al.}(2020)\citenamefont {Akimov}
  \emph {et~al.}}]{RED-100:2019rpf}%
  \BibitemOpen
  \bibfield  {author} {\bibinfo {author} {\bibfnamefont {D.~Y.}\ \bibnamefont
  {Akimov}} \emph {et~al.} (\bibinfo {collaboration} {RED-100}),\ }\href
  {\doibase 10.1088/1748-0221/15/02/P02020} {\bibfield  {journal} {\bibinfo
  {journal} {JINST}\ }\textbf {\bibinfo {volume} {15}},\ \bibinfo {pages}
  {P02020} (\bibinfo {year} {2020})},\ \Eprint
  {http://arxiv.org/abs/1910.06190} {arXiv:1910.06190 [physics.ins-det]}
  \BibitemShut {NoStop}%
\bibitem [{\citenamefont {Wei}\ and\ \citenamefont
  {Huang}(2019)}]{wei2019halide}%
  \BibitemOpen
  \bibfield  {author} {\bibinfo {author} {\bibfnamefont {H.}~\bibnamefont
  {Wei}}\ and\ \bibinfo {author} {\bibfnamefont {J.}~\bibnamefont {Huang}},\
  }\href@noop {} {\bibfield  {journal} {\bibinfo  {journal} {Nature
  communications}\ }\textbf {\bibinfo {volume} {10}},\ \bibinfo {pages} {1}
  (\bibinfo {year} {2019})}\BibitemShut {NoStop}%
\bibitem [{\citenamefont {Kojima}\ \emph {et~al.}(2009)\citenamefont {Kojima},
  \citenamefont {Teshima}, \citenamefont {Shirai},\ and\ \citenamefont
  {Miyasaka}}]{kojima2009organometal}%
  \BibitemOpen
  \bibfield  {author} {\bibinfo {author} {\bibfnamefont {A.}~\bibnamefont
  {Kojima}}, \bibinfo {author} {\bibfnamefont {K.}~\bibnamefont {Teshima}},
  \bibinfo {author} {\bibfnamefont {Y.}~\bibnamefont {Shirai}}, \ and\ \bibinfo
  {author} {\bibfnamefont {T.}~\bibnamefont {Miyasaka}},\ }\href@noop {}
  {\bibfield  {journal} {\bibinfo  {journal} {Journal of the american chemical
  society}\ }\textbf {\bibinfo {volume} {131}},\ \bibinfo {pages} {6050}
  (\bibinfo {year} {2009})}\BibitemShut {NoStop}%
\bibitem [{\citenamefont {Green}\ \emph {et~al.}(2014)\citenamefont {Green},
  \citenamefont {Ho-Baillie},\ and\ \citenamefont
  {Snaith}}]{green2014emergence}%
  \BibitemOpen
  \bibfield  {author} {\bibinfo {author} {\bibfnamefont {M.~A.}\ \bibnamefont
  {Green}}, \bibinfo {author} {\bibfnamefont {A.}~\bibnamefont {Ho-Baillie}}, \
  and\ \bibinfo {author} {\bibfnamefont {H.~J.}\ \bibnamefont {Snaith}},\
  }\href@noop {} {\bibfield  {journal} {\bibinfo  {journal} {Nature photonics}\
  }\textbf {\bibinfo {volume} {8}},\ \bibinfo {pages} {506} (\bibinfo {year}
  {2014})}\BibitemShut {NoStop}%
\bibitem [{\citenamefont {Jung}\ and\ \citenamefont
  {Park}(2015)}]{jung2015perovskite}%
  \BibitemOpen
  \bibfield  {author} {\bibinfo {author} {\bibfnamefont {H.~S.}\ \bibnamefont
  {Jung}}\ and\ \bibinfo {author} {\bibfnamefont {N.-G.}\ \bibnamefont
  {Park}},\ }\href@noop {} {\bibfield  {journal} {\bibinfo  {journal} {small}\
  }\textbf {\bibinfo {volume} {11}},\ \bibinfo {pages} {10} (\bibinfo {year}
  {2015})}\BibitemShut {NoStop}%
\bibitem [{\citenamefont {Park}(2015)}]{park2015perovskite}%
  \BibitemOpen
  \bibfield  {author} {\bibinfo {author} {\bibfnamefont {N.-G.}\ \bibnamefont
  {Park}},\ }\href@noop {} {\bibfield  {journal} {\bibinfo  {journal}
  {Materials today}\ }\textbf {\bibinfo {volume} {18}},\ \bibinfo {pages} {65}
  (\bibinfo {year} {2015})}\BibitemShut {NoStop}%
\bibitem [{\citenamefont {Correa-Baena}\ \emph {et~al.}(2017)\citenamefont
  {Correa-Baena}, \citenamefont {Saliba}, \citenamefont {Buonassisi},
  \citenamefont {Gr{\"a}tzel}, \citenamefont {Abate}, \citenamefont {Tress},\
  and\ \citenamefont {Hagfeldt}}]{correa2017promises}%
  \BibitemOpen
  \bibfield  {author} {\bibinfo {author} {\bibfnamefont {J.-P.}\ \bibnamefont
  {Correa-Baena}}, \bibinfo {author} {\bibfnamefont {M.}~\bibnamefont
  {Saliba}}, \bibinfo {author} {\bibfnamefont {T.}~\bibnamefont {Buonassisi}},
  \bibinfo {author} {\bibfnamefont {M.}~\bibnamefont {Gr{\"a}tzel}}, \bibinfo
  {author} {\bibfnamefont {A.}~\bibnamefont {Abate}}, \bibinfo {author}
  {\bibfnamefont {W.}~\bibnamefont {Tress}}, \ and\ \bibinfo {author}
  {\bibfnamefont {A.}~\bibnamefont {Hagfeldt}},\ }\href@noop {} {\bibfield
  {journal} {\bibinfo  {journal} {Science}\ }\textbf {\bibinfo {volume}
  {358}},\ \bibinfo {pages} {739} (\bibinfo {year} {2017})}\BibitemShut
  {NoStop}%
\bibitem [{\citenamefont {Huang}\ \emph {et~al.}(2017)\citenamefont {Huang},
  \citenamefont {Yuan}, \citenamefont {Shao},\ and\ \citenamefont
  {Yan}}]{huang2017understanding}%
  \BibitemOpen
  \bibfield  {author} {\bibinfo {author} {\bibfnamefont {J.}~\bibnamefont
  {Huang}}, \bibinfo {author} {\bibfnamefont {Y.}~\bibnamefont {Yuan}},
  \bibinfo {author} {\bibfnamefont {Y.}~\bibnamefont {Shao}}, \ and\ \bibinfo
  {author} {\bibfnamefont {Y.}~\bibnamefont {Yan}},\ }\href@noop {} {\bibfield
  {journal} {\bibinfo  {journal} {Nature Reviews Materials}\ }\textbf {\bibinfo
  {volume} {2}},\ \bibinfo {pages} {1} (\bibinfo {year} {2017})}\BibitemShut
  {NoStop}%
\bibitem [{\citenamefont {Yang}\ \emph {et~al.}(2018)\citenamefont {Yang},
  \citenamefont {Yang}, \citenamefont {Wang}, \citenamefont {Wu}, \citenamefont
  {Zhu}, \citenamefont {Feng}, \citenamefont {Ren}, \citenamefont {Fang},
  \citenamefont {Priya},\ and\ \citenamefont {Liu}}]{yang2018high}%
  \BibitemOpen
  \bibfield  {author} {\bibinfo {author} {\bibfnamefont {D.}~\bibnamefont
  {Yang}}, \bibinfo {author} {\bibfnamefont {R.}~\bibnamefont {Yang}}, \bibinfo
  {author} {\bibfnamefont {K.}~\bibnamefont {Wang}}, \bibinfo {author}
  {\bibfnamefont {C.}~\bibnamefont {Wu}}, \bibinfo {author} {\bibfnamefont
  {X.}~\bibnamefont {Zhu}}, \bibinfo {author} {\bibfnamefont {J.}~\bibnamefont
  {Feng}}, \bibinfo {author} {\bibfnamefont {X.}~\bibnamefont {Ren}}, \bibinfo
  {author} {\bibfnamefont {G.}~\bibnamefont {Fang}}, \bibinfo {author}
  {\bibfnamefont {S.}~\bibnamefont {Priya}}, \ and\ \bibinfo {author}
  {\bibfnamefont {S.~F.}\ \bibnamefont {Liu}},\ }\href@noop {} {\bibfield
  {journal} {\bibinfo  {journal} {Nature communications}\ }\textbf {\bibinfo
  {volume} {9}},\ \bibinfo {pages} {1} (\bibinfo {year} {2018})}\BibitemShut
  {NoStop}%
\bibitem [{\citenamefont {Kim}\ \emph {et~al.}(2020)\citenamefont {Kim},
  \citenamefont {Lee}, \citenamefont {Jung}, \citenamefont {Shin},\ and\
  \citenamefont {Park}}]{kim2020high}%
  \BibitemOpen
  \bibfield  {author} {\bibinfo {author} {\bibfnamefont {J.~Y.}\ \bibnamefont
  {Kim}}, \bibinfo {author} {\bibfnamefont {J.-W.}\ \bibnamefont {Lee}},
  \bibinfo {author} {\bibfnamefont {H.~S.}\ \bibnamefont {Jung}}, \bibinfo
  {author} {\bibfnamefont {H.}~\bibnamefont {Shin}}, \ and\ \bibinfo {author}
  {\bibfnamefont {N.-G.}\ \bibnamefont {Park}},\ }\href@noop {} {\bibfield
  {journal} {\bibinfo  {journal} {Chemical Reviews}\ }\textbf {\bibinfo
  {volume} {120}},\ \bibinfo {pages} {7867} (\bibinfo {year}
  {2020})}\BibitemShut {NoStop}%
\bibitem [{\citenamefont {Yoo}\ \emph {et~al.}(2021)\citenamefont {Yoo},
  \citenamefont {Seo}, \citenamefont {Chua}, \citenamefont {Park},
  \citenamefont {Lu}, \citenamefont {Rotermund}, \citenamefont {Kim},
  \citenamefont {Moon}, \citenamefont {Jeon}, \citenamefont {Correa-Baena}
  \emph {et~al.}}]{yoo2021efficient}%
  \BibitemOpen
  \bibfield  {author} {\bibinfo {author} {\bibfnamefont {J.~J.}\ \bibnamefont
  {Yoo}}, \bibinfo {author} {\bibfnamefont {G.}~\bibnamefont {Seo}}, \bibinfo
  {author} {\bibfnamefont {M.~R.}\ \bibnamefont {Chua}}, \bibinfo {author}
  {\bibfnamefont {T.~G.}\ \bibnamefont {Park}}, \bibinfo {author}
  {\bibfnamefont {Y.}~\bibnamefont {Lu}}, \bibinfo {author} {\bibfnamefont
  {F.}~\bibnamefont {Rotermund}}, \bibinfo {author} {\bibfnamefont {Y.-K.}\
  \bibnamefont {Kim}}, \bibinfo {author} {\bibfnamefont {C.~S.}\ \bibnamefont
  {Moon}}, \bibinfo {author} {\bibfnamefont {N.~J.}\ \bibnamefont {Jeon}},
  \bibinfo {author} {\bibfnamefont {J.-P.}\ \bibnamefont {Correa-Baena}},
  \emph {et~al.},\ }\href@noop {} {\bibfield  {journal} {\bibinfo  {journal}
  {Nature}\ }\textbf {\bibinfo {volume} {590}},\ \bibinfo {pages} {587}
  (\bibinfo {year} {2021})}\BibitemShut {NoStop}%
\bibitem [{\citenamefont {Miyata}\ \emph {et~al.}(2015)\citenamefont {Miyata},
  \citenamefont {Mitioglu}, \citenamefont {Plochocka}, \citenamefont
  {Portugall}, \citenamefont {Wang}, \citenamefont {Stranks}, \citenamefont
  {Snaith},\ and\ \citenamefont {Nicholas}}]{miyata2015direct}%
  \BibitemOpen
  \bibfield  {author} {\bibinfo {author} {\bibfnamefont {A.}~\bibnamefont
  {Miyata}}, \bibinfo {author} {\bibfnamefont {A.}~\bibnamefont {Mitioglu}},
  \bibinfo {author} {\bibfnamefont {P.}~\bibnamefont {Plochocka}}, \bibinfo
  {author} {\bibfnamefont {O.}~\bibnamefont {Portugall}}, \bibinfo {author}
  {\bibfnamefont {J.~T.-W.}\ \bibnamefont {Wang}}, \bibinfo {author}
  {\bibfnamefont {S.~D.}\ \bibnamefont {Stranks}}, \bibinfo {author}
  {\bibfnamefont {H.~J.}\ \bibnamefont {Snaith}}, \ and\ \bibinfo {author}
  {\bibfnamefont {R.~J.}\ \bibnamefont {Nicholas}},\ }\href@noop {} {\bibfield
  {journal} {\bibinfo  {journal} {Nature Physics}\ }\textbf {\bibinfo {volume}
  {11}},\ \bibinfo {pages} {582} (\bibinfo {year} {2015})}\BibitemShut
  {NoStop}%
\bibitem [{\citenamefont {Stranks}\ \emph {et~al.}(2013)\citenamefont
  {Stranks}, \citenamefont {Eperon}, \citenamefont {Grancini}, \citenamefont
  {Menelaou}, \citenamefont {Alcocer}, \citenamefont {Leijtens}, \citenamefont
  {Herz}, \citenamefont {Petrozza},\ and\ \citenamefont
  {Snaith}}]{stranks2013electron}%
  \BibitemOpen
  \bibfield  {author} {\bibinfo {author} {\bibfnamefont {S.~D.}\ \bibnamefont
  {Stranks}}, \bibinfo {author} {\bibfnamefont {G.~E.}\ \bibnamefont {Eperon}},
  \bibinfo {author} {\bibfnamefont {G.}~\bibnamefont {Grancini}}, \bibinfo
  {author} {\bibfnamefont {C.}~\bibnamefont {Menelaou}}, \bibinfo {author}
  {\bibfnamefont {M.~J.}\ \bibnamefont {Alcocer}}, \bibinfo {author}
  {\bibfnamefont {T.}~\bibnamefont {Leijtens}}, \bibinfo {author}
  {\bibfnamefont {L.~M.}\ \bibnamefont {Herz}}, \bibinfo {author}
  {\bibfnamefont {A.}~\bibnamefont {Petrozza}}, \ and\ \bibinfo {author}
  {\bibfnamefont {H.~J.}\ \bibnamefont {Snaith}},\ }\href@noop {} {\bibfield
  {journal} {\bibinfo  {journal} {Science}\ }\textbf {\bibinfo {volume}
  {342}},\ \bibinfo {pages} {341} (\bibinfo {year} {2013})}\BibitemShut
  {NoStop}%
\bibitem [{\citenamefont {Ju}\ \emph {et~al.}(2018)\citenamefont {Ju},
  \citenamefont {Dang}, \citenamefont {Zhu}, \citenamefont {Liu}, \citenamefont
  {Chueh}, \citenamefont {Li}, \citenamefont {Wang}, \citenamefont {Hu},
  \citenamefont {Jen},\ and\ \citenamefont {Tao}}]{ju2018tunable}%
  \BibitemOpen
  \bibfield  {author} {\bibinfo {author} {\bibfnamefont {D.}~\bibnamefont
  {Ju}}, \bibinfo {author} {\bibfnamefont {Y.}~\bibnamefont {Dang}}, \bibinfo
  {author} {\bibfnamefont {Z.}~\bibnamefont {Zhu}}, \bibinfo {author}
  {\bibfnamefont {H.}~\bibnamefont {Liu}}, \bibinfo {author} {\bibfnamefont
  {C.-C.}\ \bibnamefont {Chueh}}, \bibinfo {author} {\bibfnamefont
  {X.}~\bibnamefont {Li}}, \bibinfo {author} {\bibfnamefont {L.}~\bibnamefont
  {Wang}}, \bibinfo {author} {\bibfnamefont {X.}~\bibnamefont {Hu}}, \bibinfo
  {author} {\bibfnamefont {A.~K.-Y.}\ \bibnamefont {Jen}}, \ and\ \bibinfo
  {author} {\bibfnamefont {X.}~\bibnamefont {Tao}},\ }\href@noop {} {\bibfield
  {journal} {\bibinfo  {journal} {Chemistry of Materials}\ }\textbf {\bibinfo
  {volume} {30}},\ \bibinfo {pages} {1556} (\bibinfo {year}
  {2018})}\BibitemShut {NoStop}%
\bibitem [{\citenamefont {Unger}\ \emph {et~al.}(2017)\citenamefont {Unger},
  \citenamefont {Kegelmann}, \citenamefont {Suchan}, \citenamefont
  {S{\"o}rell}, \citenamefont {Korte},\ and\ \citenamefont
  {Albrecht}}]{unger2017roadmap}%
  \BibitemOpen
  \bibfield  {author} {\bibinfo {author} {\bibfnamefont {E.}~\bibnamefont
  {Unger}}, \bibinfo {author} {\bibfnamefont {L.}~\bibnamefont {Kegelmann}},
  \bibinfo {author} {\bibfnamefont {K.}~\bibnamefont {Suchan}}, \bibinfo
  {author} {\bibfnamefont {D.}~\bibnamefont {S{\"o}rell}}, \bibinfo {author}
  {\bibfnamefont {L.}~\bibnamefont {Korte}}, \ and\ \bibinfo {author}
  {\bibfnamefont {S.}~\bibnamefont {Albrecht}},\ }\href@noop {} {\bibfield
  {journal} {\bibinfo  {journal} {Journal of Materials Chemistry A}\ }\textbf
  {\bibinfo {volume} {5}},\ \bibinfo {pages} {11401} (\bibinfo {year}
  {2017})}\BibitemShut {NoStop}%
\bibitem [{\citenamefont {Pisoni}\ \emph {et~al.}(2014)\citenamefont {Pisoni},
  \citenamefont {Jacimovic}, \citenamefont {Barisic}, \citenamefont {Spina},
  \citenamefont {Ga{\'a}l}, \citenamefont {Forr{\'o}},\ and\ \citenamefont
  {Horv{\'a}th}}]{pisoni2014ultra}%
  \BibitemOpen
  \bibfield  {author} {\bibinfo {author} {\bibfnamefont {A.}~\bibnamefont
  {Pisoni}}, \bibinfo {author} {\bibfnamefont {J.}~\bibnamefont {Jacimovic}},
  \bibinfo {author} {\bibfnamefont {O.~S.}\ \bibnamefont {Barisic}}, \bibinfo
  {author} {\bibfnamefont {M.}~\bibnamefont {Spina}}, \bibinfo {author}
  {\bibfnamefont {R.}~\bibnamefont {Ga{\'a}l}}, \bibinfo {author}
  {\bibfnamefont {L.}~\bibnamefont {Forr{\'o}}}, \ and\ \bibinfo {author}
  {\bibfnamefont {E.}~\bibnamefont {Horv{\'a}th}},\ }\href@noop {} {\bibfield
  {journal} {\bibinfo  {journal} {The journal of physical chemistry letters}\
  }\textbf {\bibinfo {volume} {5}},\ \bibinfo {pages} {2488} (\bibinfo {year}
  {2014})}\BibitemShut {NoStop}%
\bibitem [{\citenamefont {Emara}\ \emph {et~al.}(2016)\citenamefont {Emara},
  \citenamefont {Schnier}, \citenamefont {Pourdavoud}, \citenamefont {Riedl},
  \citenamefont {Meerholz},\ and\ \citenamefont {Olthof}}]{emara2016impact}%
  \BibitemOpen
  \bibfield  {author} {\bibinfo {author} {\bibfnamefont {J.}~\bibnamefont
  {Emara}}, \bibinfo {author} {\bibfnamefont {T.}~\bibnamefont {Schnier}},
  \bibinfo {author} {\bibfnamefont {N.}~\bibnamefont {Pourdavoud}}, \bibinfo
  {author} {\bibfnamefont {T.}~\bibnamefont {Riedl}}, \bibinfo {author}
  {\bibfnamefont {K.}~\bibnamefont {Meerholz}}, \ and\ \bibinfo {author}
  {\bibfnamefont {S.}~\bibnamefont {Olthof}},\ }\href@noop {} {\bibfield
  {journal} {\bibinfo  {journal} {Advanced Materials}\ }\textbf {\bibinfo
  {volume} {28}},\ \bibinfo {pages} {553} (\bibinfo {year} {2016})}\BibitemShut
  {NoStop}%
\bibitem [{\citenamefont {Xiao}\ \emph {et~al.}(2021)\citenamefont {Xiao},
  \citenamefont {Jia}, \citenamefont {Bu}, \citenamefont {Li}, \citenamefont
  {Liu}, \citenamefont {Liu}, \citenamefont {Yang},\ and\ \citenamefont
  {Liu}}]{xiao2021grain}%
  \BibitemOpen
  \bibfield  {author} {\bibinfo {author} {\bibfnamefont {Y.}~\bibnamefont
  {Xiao}}, \bibinfo {author} {\bibfnamefont {S.}~\bibnamefont {Jia}}, \bibinfo
  {author} {\bibfnamefont {N.}~\bibnamefont {Bu}}, \bibinfo {author}
  {\bibfnamefont {N.}~\bibnamefont {Li}}, \bibinfo {author} {\bibfnamefont
  {Y.}~\bibnamefont {Liu}}, \bibinfo {author} {\bibfnamefont {M.}~\bibnamefont
  {Liu}}, \bibinfo {author} {\bibfnamefont {Z.}~\bibnamefont {Yang}}, \ and\
  \bibinfo {author} {\bibfnamefont {S.~F.}\ \bibnamefont {Liu}},\ }\href@noop
  {} {\bibfield  {journal} {\bibinfo  {journal} {Journal of Materials Chemistry
  A}\ }\textbf {\bibinfo {volume} {9}},\ \bibinfo {pages} {25603} (\bibinfo
  {year} {2021})}\BibitemShut {NoStop}%
\bibitem [{\citenamefont {Shrestha}\ \emph {et~al.}(2017)\citenamefont
  {Shrestha}, \citenamefont {Fischer}, \citenamefont {Matt}, \citenamefont
  {Feldner}, \citenamefont {Michel}, \citenamefont {Osvet}, \citenamefont
  {Levchuk}, \citenamefont {Merle}, \citenamefont {Golkar}, \citenamefont
  {Chen} \emph {et~al.}}]{shrestha2017high}%
  \BibitemOpen
  \bibfield  {author} {\bibinfo {author} {\bibfnamefont {S.}~\bibnamefont
  {Shrestha}}, \bibinfo {author} {\bibfnamefont {R.}~\bibnamefont {Fischer}},
  \bibinfo {author} {\bibfnamefont {G.~J.}\ \bibnamefont {Matt}}, \bibinfo
  {author} {\bibfnamefont {P.}~\bibnamefont {Feldner}}, \bibinfo {author}
  {\bibfnamefont {T.}~\bibnamefont {Michel}}, \bibinfo {author} {\bibfnamefont
  {A.}~\bibnamefont {Osvet}}, \bibinfo {author} {\bibfnamefont
  {I.}~\bibnamefont {Levchuk}}, \bibinfo {author} {\bibfnamefont
  {B.}~\bibnamefont {Merle}}, \bibinfo {author} {\bibfnamefont
  {S.}~\bibnamefont {Golkar}}, \bibinfo {author} {\bibfnamefont
  {H.}~\bibnamefont {Chen}},  \emph {et~al.},\ }\href@noop {} {\bibfield
  {journal} {\bibinfo  {journal} {Nature Photonics}\ }\textbf {\bibinfo
  {volume} {11}},\ \bibinfo {pages} {436} (\bibinfo {year} {2017})}\BibitemShut
  {NoStop}%
\bibitem [{\citenamefont {Wei}\ \emph {et~al.}(2017{\natexlab{a}})\citenamefont
  {Wei}, \citenamefont {Zhang}, \citenamefont {Xu}, \citenamefont {Wei},
  \citenamefont {Fang}, \citenamefont {Wang}, \citenamefont {Deng},
  \citenamefont {Li}, \citenamefont {Gruverman}, \citenamefont {Cao} \emph
  {et~al.}}]{wei2017monolithic}%
  \BibitemOpen
  \bibfield  {author} {\bibinfo {author} {\bibfnamefont {W.}~\bibnamefont
  {Wei}}, \bibinfo {author} {\bibfnamefont {Y.}~\bibnamefont {Zhang}}, \bibinfo
  {author} {\bibfnamefont {Q.}~\bibnamefont {Xu}}, \bibinfo {author}
  {\bibfnamefont {H.}~\bibnamefont {Wei}}, \bibinfo {author} {\bibfnamefont
  {Y.}~\bibnamefont {Fang}}, \bibinfo {author} {\bibfnamefont {Q.}~\bibnamefont
  {Wang}}, \bibinfo {author} {\bibfnamefont {Y.}~\bibnamefont {Deng}}, \bibinfo
  {author} {\bibfnamefont {T.}~\bibnamefont {Li}}, \bibinfo {author}
  {\bibfnamefont {A.}~\bibnamefont {Gruverman}}, \bibinfo {author}
  {\bibfnamefont {L.}~\bibnamefont {Cao}},  \emph {et~al.},\ }\href@noop {}
  {\bibfield  {journal} {\bibinfo  {journal} {Nature Photonics}\ }\textbf
  {\bibinfo {volume} {11}},\ \bibinfo {pages} {315} (\bibinfo {year}
  {2017}{\natexlab{a}})}\BibitemShut {NoStop}%
\bibitem [{\citenamefont {Kim}\ \emph {et~al.}(2017)\citenamefont {Kim},
  \citenamefont {Kim}, \citenamefont {Son}, \citenamefont {Jeong},
  \citenamefont {Seo}, \citenamefont {Choi}, \citenamefont {Han}, \citenamefont
  {Lee},\ and\ \citenamefont {Park}}]{kim2017printable}%
  \BibitemOpen
  \bibfield  {author} {\bibinfo {author} {\bibfnamefont {Y.~C.}\ \bibnamefont
  {Kim}}, \bibinfo {author} {\bibfnamefont {K.~H.}\ \bibnamefont {Kim}},
  \bibinfo {author} {\bibfnamefont {D.-Y.}\ \bibnamefont {Son}}, \bibinfo
  {author} {\bibfnamefont {D.-N.}\ \bibnamefont {Jeong}}, \bibinfo {author}
  {\bibfnamefont {J.-Y.}\ \bibnamefont {Seo}}, \bibinfo {author} {\bibfnamefont
  {Y.~S.}\ \bibnamefont {Choi}}, \bibinfo {author} {\bibfnamefont {I.~T.}\
  \bibnamefont {Han}}, \bibinfo {author} {\bibfnamefont {S.~Y.}\ \bibnamefont
  {Lee}}, \ and\ \bibinfo {author} {\bibfnamefont {N.-G.}\ \bibnamefont
  {Park}},\ }\href@noop {} {\bibfield  {journal} {\bibinfo  {journal} {Nature}\
  }\textbf {\bibinfo {volume} {550}},\ \bibinfo {pages} {87} (\bibinfo {year}
  {2017})}\BibitemShut {NoStop}%
\bibitem [{\citenamefont {Garc{\'\i}a~de Arquer}\ \emph
  {et~al.}(2017)\citenamefont {Garc{\'\i}a~de Arquer}, \citenamefont {Armin},
  \citenamefont {Meredith},\ and\ \citenamefont
  {Sargent}}]{garcia2017solution}%
  \BibitemOpen
  \bibfield  {author} {\bibinfo {author} {\bibfnamefont {F.~P.}\ \bibnamefont
  {Garc{\'\i}a~de Arquer}}, \bibinfo {author} {\bibfnamefont {A.}~\bibnamefont
  {Armin}}, \bibinfo {author} {\bibfnamefont {P.}~\bibnamefont {Meredith}}, \
  and\ \bibinfo {author} {\bibfnamefont {E.~H.}\ \bibnamefont {Sargent}},\
  }\href@noop {} {\bibfield  {journal} {\bibinfo  {journal} {Nature Reviews
  Materials}\ }\textbf {\bibinfo {volume} {2}},\ \bibinfo {pages} {1} (\bibinfo
  {year} {2017})}\BibitemShut {NoStop}%
\bibitem [{\citenamefont {Gill}\ \emph {et~al.}(2018)\citenamefont {Gill},
  \citenamefont {Elshahat}, \citenamefont {Kokil}, \citenamefont {Li},
  \citenamefont {Mosurkal}, \citenamefont {Zygmanski}, \citenamefont {Sajo},\
  and\ \citenamefont {Kumar}}]{gill2018flexible}%
  \BibitemOpen
  \bibfield  {author} {\bibinfo {author} {\bibfnamefont {H.~S.}\ \bibnamefont
  {Gill}}, \bibinfo {author} {\bibfnamefont {B.}~\bibnamefont {Elshahat}},
  \bibinfo {author} {\bibfnamefont {A.}~\bibnamefont {Kokil}}, \bibinfo
  {author} {\bibfnamefont {L.}~\bibnamefont {Li}}, \bibinfo {author}
  {\bibfnamefont {R.}~\bibnamefont {Mosurkal}}, \bibinfo {author}
  {\bibfnamefont {P.}~\bibnamefont {Zygmanski}}, \bibinfo {author}
  {\bibfnamefont {E.}~\bibnamefont {Sajo}}, \ and\ \bibinfo {author}
  {\bibfnamefont {J.}~\bibnamefont {Kumar}},\ }\href@noop {} {\bibfield
  {journal} {\bibinfo  {journal} {Physics in Medicine}\ }\textbf {\bibinfo
  {volume} {5}},\ \bibinfo {pages} {20} (\bibinfo {year} {2018})}\BibitemShut
  {NoStop}%
\bibitem [{\citenamefont {Zhuang}\ \emph {et~al.}(2019)\citenamefont {Zhuang},
  \citenamefont {Wang}, \citenamefont {Ma}, \citenamefont {Wu}, \citenamefont
  {Chen}, \citenamefont {Tang}, \citenamefont {Zhu}, \citenamefont {Liu},
  \citenamefont {Wu}, \citenamefont {Zhou} \emph {et~al.}}]{zhuang2019highly}%
  \BibitemOpen
  \bibfield  {author} {\bibinfo {author} {\bibfnamefont {R.}~\bibnamefont
  {Zhuang}}, \bibinfo {author} {\bibfnamefont {X.}~\bibnamefont {Wang}},
  \bibinfo {author} {\bibfnamefont {W.}~\bibnamefont {Ma}}, \bibinfo {author}
  {\bibfnamefont {Y.}~\bibnamefont {Wu}}, \bibinfo {author} {\bibfnamefont
  {X.}~\bibnamefont {Chen}}, \bibinfo {author} {\bibfnamefont {L.}~\bibnamefont
  {Tang}}, \bibinfo {author} {\bibfnamefont {H.}~\bibnamefont {Zhu}}, \bibinfo
  {author} {\bibfnamefont {J.}~\bibnamefont {Liu}}, \bibinfo {author}
  {\bibfnamefont {L.}~\bibnamefont {Wu}}, \bibinfo {author} {\bibfnamefont
  {W.}~\bibnamefont {Zhou}},  \emph {et~al.},\ }\href@noop {} {\bibfield
  {journal} {\bibinfo  {journal} {Nature Photonics}\ }\textbf {\bibinfo
  {volume} {13}},\ \bibinfo {pages} {602} (\bibinfo {year} {2019})}\BibitemShut
  {NoStop}%
\bibitem [{\citenamefont {Zhou}\ \emph {et~al.}(2020)\citenamefont {Zhou},
  \citenamefont {Li}, \citenamefont {Lan}, \citenamefont {Wang}, \citenamefont
  {Ding},\ and\ \citenamefont {Jin}}]{zhou2020halide}%
  \BibitemOpen
  \bibfield  {author} {\bibinfo {author} {\bibfnamefont {F.}~\bibnamefont
  {Zhou}}, \bibinfo {author} {\bibfnamefont {Z.}~\bibnamefont {Li}}, \bibinfo
  {author} {\bibfnamefont {W.}~\bibnamefont {Lan}}, \bibinfo {author}
  {\bibfnamefont {Q.}~\bibnamefont {Wang}}, \bibinfo {author} {\bibfnamefont
  {L.}~\bibnamefont {Ding}}, \ and\ \bibinfo {author} {\bibfnamefont
  {Z.}~\bibnamefont {Jin}},\ }\href@noop {} {\bibfield  {journal} {\bibinfo
  {journal} {Small Methods}\ }\textbf {\bibinfo {volume} {4}},\ \bibinfo
  {pages} {2000506} (\bibinfo {year} {2020})}\BibitemShut {NoStop}%
\bibitem [{\citenamefont {Li}\ \emph {et~al.}(2020)\citenamefont {Li},
  \citenamefont {Meng}, \citenamefont {Huang}, \citenamefont {Yang},
  \citenamefont {Xu},\ and\ \citenamefont {Zeng}}]{li2020all}%
  \BibitemOpen
  \bibfield  {author} {\bibinfo {author} {\bibfnamefont {X.}~\bibnamefont
  {Li}}, \bibinfo {author} {\bibfnamefont {C.}~\bibnamefont {Meng}}, \bibinfo
  {author} {\bibfnamefont {B.}~\bibnamefont {Huang}}, \bibinfo {author}
  {\bibfnamefont {D.}~\bibnamefont {Yang}}, \bibinfo {author} {\bibfnamefont
  {X.}~\bibnamefont {Xu}}, \ and\ \bibinfo {author} {\bibfnamefont
  {H.}~\bibnamefont {Zeng}},\ }\href@noop {} {\bibfield  {journal} {\bibinfo
  {journal} {Advanced Optical Materials}\ }\textbf {\bibinfo {volume} {8}},\
  \bibinfo {pages} {2000273} (\bibinfo {year} {2020})}\BibitemShut {NoStop}%
\bibitem [{\citenamefont {Su}\ \emph {et~al.}(2020)\citenamefont {Su},
  \citenamefont {Ma},\ and\ \citenamefont {Yang}}]{su2020perovskite}%
  \BibitemOpen
  \bibfield  {author} {\bibinfo {author} {\bibfnamefont {Y.}~\bibnamefont
  {Su}}, \bibinfo {author} {\bibfnamefont {W.}~\bibnamefont {Ma}}, \ and\
  \bibinfo {author} {\bibfnamefont {Y.~M.}\ \bibnamefont {Yang}},\ }\href@noop
  {} {\bibfield  {journal} {\bibinfo  {journal} {Journal of Semiconductors}\
  }\textbf {\bibinfo {volume} {41}},\ \bibinfo {pages} {051204} (\bibinfo
  {year} {2020})}\BibitemShut {NoStop}%
\bibitem [{\citenamefont {Dong}\ \emph {et~al.}(2015)\citenamefont {Dong},
  \citenamefont {Fang}, \citenamefont {Shao}, \citenamefont {Mulligan},
  \citenamefont {Qiu}, \citenamefont {Cao},\ and\ \citenamefont
  {Huang}}]{dong2015electron}%
  \BibitemOpen
  \bibfield  {author} {\bibinfo {author} {\bibfnamefont {Q.}~\bibnamefont
  {Dong}}, \bibinfo {author} {\bibfnamefont {Y.}~\bibnamefont {Fang}}, \bibinfo
  {author} {\bibfnamefont {Y.}~\bibnamefont {Shao}}, \bibinfo {author}
  {\bibfnamefont {P.}~\bibnamefont {Mulligan}}, \bibinfo {author}
  {\bibfnamefont {J.}~\bibnamefont {Qiu}}, \bibinfo {author} {\bibfnamefont
  {L.}~\bibnamefont {Cao}}, \ and\ \bibinfo {author} {\bibfnamefont
  {J.}~\bibnamefont {Huang}},\ }\href@noop {} {\bibfield  {journal} {\bibinfo
  {journal} {Science}\ }\textbf {\bibinfo {volume} {347}},\ \bibinfo {pages}
  {967} (\bibinfo {year} {2015})}\BibitemShut {NoStop}%
\bibitem [{\citenamefont {Yakunin}\ \emph {et~al.}(2015)\citenamefont
  {Yakunin}, \citenamefont {Sytnyk}, \citenamefont {Kriegner}, \citenamefont
  {Shrestha}, \citenamefont {Richter}, \citenamefont {Matt}, \citenamefont
  {Azimi}, \citenamefont {Brabec}, \citenamefont {Stangl}, \citenamefont
  {Kovalenko} \emph {et~al.}}]{yakunin2015detection}%
  \BibitemOpen
  \bibfield  {author} {\bibinfo {author} {\bibfnamefont {S.}~\bibnamefont
  {Yakunin}}, \bibinfo {author} {\bibfnamefont {M.}~\bibnamefont {Sytnyk}},
  \bibinfo {author} {\bibfnamefont {D.}~\bibnamefont {Kriegner}}, \bibinfo
  {author} {\bibfnamefont {S.}~\bibnamefont {Shrestha}}, \bibinfo {author}
  {\bibfnamefont {M.}~\bibnamefont {Richter}}, \bibinfo {author} {\bibfnamefont
  {G.~J.}\ \bibnamefont {Matt}}, \bibinfo {author} {\bibfnamefont
  {H.}~\bibnamefont {Azimi}}, \bibinfo {author} {\bibfnamefont {C.~J.}\
  \bibnamefont {Brabec}}, \bibinfo {author} {\bibfnamefont {J.}~\bibnamefont
  {Stangl}}, \bibinfo {author} {\bibfnamefont {M.~V.}\ \bibnamefont
  {Kovalenko}},  \emph {et~al.},\ }\href@noop {} {\bibfield  {journal}
  {\bibinfo  {journal} {Nature photonics}\ }\textbf {\bibinfo {volume} {9}},\
  \bibinfo {pages} {444} (\bibinfo {year} {2015})}\BibitemShut {NoStop}%
\bibitem [{\citenamefont {Wei}\ \emph {et~al.}(2016)\citenamefont {Wei},
  \citenamefont {Fang}, \citenamefont {Mulligan}, \citenamefont {Chuirazzi},
  \citenamefont {Fang}, \citenamefont {Wang}, \citenamefont {Ecker},
  \citenamefont {Gao}, \citenamefont {Loi}, \citenamefont {Cao} \emph
  {et~al.}}]{wei2016sensitive}%
  \BibitemOpen
  \bibfield  {author} {\bibinfo {author} {\bibfnamefont {H.}~\bibnamefont
  {Wei}}, \bibinfo {author} {\bibfnamefont {Y.}~\bibnamefont {Fang}}, \bibinfo
  {author} {\bibfnamefont {P.}~\bibnamefont {Mulligan}}, \bibinfo {author}
  {\bibfnamefont {W.}~\bibnamefont {Chuirazzi}}, \bibinfo {author}
  {\bibfnamefont {H.-H.}\ \bibnamefont {Fang}}, \bibinfo {author}
  {\bibfnamefont {C.}~\bibnamefont {Wang}}, \bibinfo {author} {\bibfnamefont
  {B.~R.}\ \bibnamefont {Ecker}}, \bibinfo {author} {\bibfnamefont
  {Y.}~\bibnamefont {Gao}}, \bibinfo {author} {\bibfnamefont {M.~A.}\
  \bibnamefont {Loi}}, \bibinfo {author} {\bibfnamefont {L.}~\bibnamefont
  {Cao}},  \emph {et~al.},\ }\href@noop {} {\bibfield  {journal} {\bibinfo
  {journal} {Nature Photonics}\ }\textbf {\bibinfo {volume} {10}},\ \bibinfo
  {pages} {333} (\bibinfo {year} {2016})}\BibitemShut {NoStop}%
\bibitem [{\citenamefont {Wei}\ \emph {et~al.}(2017{\natexlab{b}})\citenamefont
  {Wei}, \citenamefont {DeSantis}, \citenamefont {Wei}, \citenamefont {Deng},
  \citenamefont {Guo}, \citenamefont {Savenije}, \citenamefont {Cao},\ and\
  \citenamefont {Huang}}]{wei2017dopant}%
  \BibitemOpen
  \bibfield  {author} {\bibinfo {author} {\bibfnamefont {H.}~\bibnamefont
  {Wei}}, \bibinfo {author} {\bibfnamefont {D.}~\bibnamefont {DeSantis}},
  \bibinfo {author} {\bibfnamefont {W.}~\bibnamefont {Wei}}, \bibinfo {author}
  {\bibfnamefont {Y.}~\bibnamefont {Deng}}, \bibinfo {author} {\bibfnamefont
  {D.}~\bibnamefont {Guo}}, \bibinfo {author} {\bibfnamefont {T.~J.}\
  \bibnamefont {Savenije}}, \bibinfo {author} {\bibfnamefont {L.}~\bibnamefont
  {Cao}}, \ and\ \bibinfo {author} {\bibfnamefont {J.}~\bibnamefont {Huang}},\
  }\href@noop {} {\bibfield  {journal} {\bibinfo  {journal} {Nature materials}\
  }\textbf {\bibinfo {volume} {16}},\ \bibinfo {pages} {826} (\bibinfo {year}
  {2017}{\natexlab{b}})}\BibitemShut {NoStop}%
\bibitem [{\citenamefont {He}\ \emph {et~al.}(2018)\citenamefont {He},
  \citenamefont {Matei}, \citenamefont {Jung}, \citenamefont {McCall},
  \citenamefont {Chen}, \citenamefont {Stoumpos}, \citenamefont {Liu},
  \citenamefont {Peters}, \citenamefont {Chung}, \citenamefont {Wessels} \emph
  {et~al.}}]{he2018high}%
  \BibitemOpen
  \bibfield  {author} {\bibinfo {author} {\bibfnamefont {Y.}~\bibnamefont
  {He}}, \bibinfo {author} {\bibfnamefont {L.}~\bibnamefont {Matei}}, \bibinfo
  {author} {\bibfnamefont {H.~J.}\ \bibnamefont {Jung}}, \bibinfo {author}
  {\bibfnamefont {K.~M.}\ \bibnamefont {McCall}}, \bibinfo {author}
  {\bibfnamefont {M.}~\bibnamefont {Chen}}, \bibinfo {author} {\bibfnamefont
  {C.~C.}\ \bibnamefont {Stoumpos}}, \bibinfo {author} {\bibfnamefont
  {Z.}~\bibnamefont {Liu}}, \bibinfo {author} {\bibfnamefont {J.~A.}\
  \bibnamefont {Peters}}, \bibinfo {author} {\bibfnamefont {D.~Y.}\
  \bibnamefont {Chung}}, \bibinfo {author} {\bibfnamefont {B.~W.}\ \bibnamefont
  {Wessels}},  \emph {et~al.},\ }\href@noop {} {\bibfield  {journal} {\bibinfo
  {journal} {Nature communications}\ }\textbf {\bibinfo {volume} {9}},\
  \bibinfo {pages} {1} (\bibinfo {year} {2018})}\BibitemShut {NoStop}%
\bibitem [{\citenamefont {He}\ \emph {et~al.}(2021)\citenamefont {He},
  \citenamefont {Petryk}, \citenamefont {Liu}, \citenamefont {Chica},
  \citenamefont {Hadar}, \citenamefont {Leak}, \citenamefont {Ke},
  \citenamefont {Spanopoulos}, \citenamefont {Lin}, \citenamefont {Chung} \emph
  {et~al.}}]{he2021cspbbr3}%
  \BibitemOpen
  \bibfield  {author} {\bibinfo {author} {\bibfnamefont {Y.}~\bibnamefont
  {He}}, \bibinfo {author} {\bibfnamefont {M.}~\bibnamefont {Petryk}}, \bibinfo
  {author} {\bibfnamefont {Z.}~\bibnamefont {Liu}}, \bibinfo {author}
  {\bibfnamefont {D.~G.}\ \bibnamefont {Chica}}, \bibinfo {author}
  {\bibfnamefont {I.}~\bibnamefont {Hadar}}, \bibinfo {author} {\bibfnamefont
  {C.}~\bibnamefont {Leak}}, \bibinfo {author} {\bibfnamefont {W.}~\bibnamefont
  {Ke}}, \bibinfo {author} {\bibfnamefont {I.}~\bibnamefont {Spanopoulos}},
  \bibinfo {author} {\bibfnamefont {W.}~\bibnamefont {Lin}}, \bibinfo {author}
  {\bibfnamefont {D.~Y.}\ \bibnamefont {Chung}},  \emph {et~al.},\ }\href@noop
  {} {\bibfield  {journal} {\bibinfo  {journal} {Nature Photonics}\ }\textbf
  {\bibinfo {volume} {15}},\ \bibinfo {pages} {36} (\bibinfo {year}
  {2021})}\BibitemShut {NoStop}%
\bibitem [{\citenamefont {Jiang}\ \emph {et~al.}(2022)\citenamefont {Jiang},
  \citenamefont {Xiong}, \citenamefont {Fan}, \citenamefont {Bao},
  \citenamefont {Xin}, \citenamefont {Pan}, \citenamefont {Fei}, \citenamefont
  {Huang}, \citenamefont {Zhou}, \citenamefont {Yao} \emph
  {et~al.}}]{jiang2022synergistic}%
  \BibitemOpen
  \bibfield  {author} {\bibinfo {author} {\bibfnamefont {J.}~\bibnamefont
  {Jiang}}, \bibinfo {author} {\bibfnamefont {M.}~\bibnamefont {Xiong}},
  \bibinfo {author} {\bibfnamefont {K.}~\bibnamefont {Fan}}, \bibinfo {author}
  {\bibfnamefont {C.}~\bibnamefont {Bao}}, \bibinfo {author} {\bibfnamefont
  {D.}~\bibnamefont {Xin}}, \bibinfo {author} {\bibfnamefont {Z.}~\bibnamefont
  {Pan}}, \bibinfo {author} {\bibfnamefont {L.}~\bibnamefont {Fei}}, \bibinfo
  {author} {\bibfnamefont {H.}~\bibnamefont {Huang}}, \bibinfo {author}
  {\bibfnamefont {L.}~\bibnamefont {Zhou}}, \bibinfo {author} {\bibfnamefont
  {K.}~\bibnamefont {Yao}},  \emph {et~al.},\ }\href@noop {} {\bibfield
  {journal} {\bibinfo  {journal} {Nature Photonics}\ }\textbf {\bibinfo
  {volume} {16}},\ \bibinfo {pages} {575} (\bibinfo {year} {2022})}\BibitemShut
  {NoStop}%
\bibitem [{\citenamefont {He}\ \emph {et~al.}(2022)\citenamefont {He},
  \citenamefont {Hadar}, \citenamefont {De~Siena}, \citenamefont {Klepov},
  \citenamefont {Pan}, \citenamefont {Chung},\ and\ \citenamefont
  {Kanatzidis}}]{he2022sensitivity}%
  \BibitemOpen
  \bibfield  {author} {\bibinfo {author} {\bibfnamefont {Y.}~\bibnamefont
  {He}}, \bibinfo {author} {\bibfnamefont {I.}~\bibnamefont {Hadar}}, \bibinfo
  {author} {\bibfnamefont {M.~C.}\ \bibnamefont {De~Siena}}, \bibinfo {author}
  {\bibfnamefont {V.~V.}\ \bibnamefont {Klepov}}, \bibinfo {author}
  {\bibfnamefont {L.}~\bibnamefont {Pan}}, \bibinfo {author} {\bibfnamefont
  {D.~Y.}\ \bibnamefont {Chung}}, \ and\ \bibinfo {author} {\bibfnamefont
  {M.~G.}\ \bibnamefont {Kanatzidis}},\ }\href@noop {} {\bibfield  {journal}
  {\bibinfo  {journal} {Advanced Functional Materials}\ ,\ \bibinfo {pages}
  {2112925}} (\bibinfo {year} {2022})}\BibitemShut {NoStop}%
\bibitem [{\citenamefont {Xie}\ \emph {et~al.}(2020)\citenamefont {Xie},
  \citenamefont {Hettiarachchi}, \citenamefont {Maddalena}, \citenamefont
  {Witkowski}, \citenamefont {Makowski}, \citenamefont {Drozdowski},
  \citenamefont {Arramel}, \citenamefont {Wee}, \citenamefont {Springham},
  \citenamefont {Vuong} \emph {et~al.}}]{xie2020lithium}%
  \BibitemOpen
  \bibfield  {author} {\bibinfo {author} {\bibfnamefont {A.}~\bibnamefont
  {Xie}}, \bibinfo {author} {\bibfnamefont {C.}~\bibnamefont {Hettiarachchi}},
  \bibinfo {author} {\bibfnamefont {F.}~\bibnamefont {Maddalena}}, \bibinfo
  {author} {\bibfnamefont {M.~E.}\ \bibnamefont {Witkowski}}, \bibinfo {author}
  {\bibfnamefont {M.}~\bibnamefont {Makowski}}, \bibinfo {author}
  {\bibfnamefont {W.}~\bibnamefont {Drozdowski}}, \bibinfo {author}
  {\bibfnamefont {A.}~\bibnamefont {Arramel}}, \bibinfo {author} {\bibfnamefont
  {A.~T.}\ \bibnamefont {Wee}}, \bibinfo {author} {\bibfnamefont {S.~V.}\
  \bibnamefont {Springham}}, \bibinfo {author} {\bibfnamefont {P.~Q.}\
  \bibnamefont {Vuong}},  \emph {et~al.},\ }\href@noop {} {\bibfield  {journal}
  {\bibinfo  {journal} {Communications Materials}\ }\textbf {\bibinfo {volume}
  {1}},\ \bibinfo {pages} {37} (\bibinfo {year} {2020})}\BibitemShut {NoStop}%
\bibitem [{\citenamefont {Yu}\ \emph {et~al.}(2020)\citenamefont {Yu},
  \citenamefont {Wang}, \citenamefont {Cao}, \citenamefont {Gu}, \citenamefont
  {Liu}, \citenamefont {Han}, \citenamefont {Huang}, \citenamefont {Zou},
  \citenamefont {Xu},\ and\ \citenamefont {Zeng}}]{yu2020two}%
  \BibitemOpen
  \bibfield  {author} {\bibinfo {author} {\bibfnamefont {D.}~\bibnamefont
  {Yu}}, \bibinfo {author} {\bibfnamefont {P.}~\bibnamefont {Wang}}, \bibinfo
  {author} {\bibfnamefont {F.}~\bibnamefont {Cao}}, \bibinfo {author}
  {\bibfnamefont {Y.}~\bibnamefont {Gu}}, \bibinfo {author} {\bibfnamefont
  {J.}~\bibnamefont {Liu}}, \bibinfo {author} {\bibfnamefont {Z.}~\bibnamefont
  {Han}}, \bibinfo {author} {\bibfnamefont {B.}~\bibnamefont {Huang}}, \bibinfo
  {author} {\bibfnamefont {Y.}~\bibnamefont {Zou}}, \bibinfo {author}
  {\bibfnamefont {X.}~\bibnamefont {Xu}}, \ and\ \bibinfo {author}
  {\bibfnamefont {H.}~\bibnamefont {Zeng}},\ }\href@noop {} {\bibfield
  {journal} {\bibinfo  {journal} {Nature Communications}\ }\textbf {\bibinfo
  {volume} {11}},\ \bibinfo {pages} {3395} (\bibinfo {year}
  {2020})}\BibitemShut {NoStop}%
\bibitem [{\citenamefont {Andri{\v{c}}evi{\'c}}\ \emph
  {et~al.}(2021)\citenamefont {Andri{\v{c}}evi{\'c}}, \citenamefont
  {N{\'a}fr{\'a}di}, \citenamefont {Koll{\'a}r}, \citenamefont
  {N{\'a}fr{\'a}di}, \citenamefont {Lilley}, \citenamefont {Kinane},
  \citenamefont {Frajtag}, \citenamefont {Sienkiewicz}, \citenamefont {Pautz},
  \citenamefont {Horv{\'a}th} \emph {et~al.}}]{andrivcevic2021hybrid}%
  \BibitemOpen
  \bibfield  {author} {\bibinfo {author} {\bibfnamefont {P.}~\bibnamefont
  {Andri{\v{c}}evi{\'c}}}, \bibinfo {author} {\bibfnamefont {G.}~\bibnamefont
  {N{\'a}fr{\'a}di}}, \bibinfo {author} {\bibfnamefont {M.}~\bibnamefont
  {Koll{\'a}r}}, \bibinfo {author} {\bibfnamefont {B.}~\bibnamefont
  {N{\'a}fr{\'a}di}}, \bibinfo {author} {\bibfnamefont {S.}~\bibnamefont
  {Lilley}}, \bibinfo {author} {\bibfnamefont {C.}~\bibnamefont {Kinane}},
  \bibinfo {author} {\bibfnamefont {P.}~\bibnamefont {Frajtag}}, \bibinfo
  {author} {\bibfnamefont {A.}~\bibnamefont {Sienkiewicz}}, \bibinfo {author}
  {\bibfnamefont {A.}~\bibnamefont {Pautz}}, \bibinfo {author} {\bibfnamefont
  {E.}~\bibnamefont {Horv{\'a}th}},  \emph {et~al.},\ }\href@noop {} {\bibfield
   {journal} {\bibinfo  {journal} {Scientific Reports}\ }\textbf {\bibinfo
  {volume} {11}},\ \bibinfo {pages} {1} (\bibinfo {year} {2021})}\BibitemShut
  {NoStop}%
\bibitem [{\citenamefont {Martineau}\ \emph {et~al.}(2004)\citenamefont
  {Martineau} \emph {et~al.}}]{EDELWEISS:2003omv}%
  \BibitemOpen
  \bibfield  {author} {\bibinfo {author} {\bibfnamefont {O.}~\bibnamefont
  {Martineau}} \emph {et~al.} (\bibinfo {collaboration} {EDELWEISS}),\ }\href
  {\doibase 10.1016/j.nima.2004.04.218} {\bibfield  {journal} {\bibinfo
  {journal} {Nucl. Instrum. Meth. A}\ }\textbf {\bibinfo {volume} {530}},\
  \bibinfo {pages} {426} (\bibinfo {year} {2004})},\ \Eprint
  {http://arxiv.org/abs/astro-ph/0310657} {arXiv:astro-ph/0310657} \BibitemShut
  {NoStop}%
\bibitem [{\citenamefont {Beeman}\ \emph {et~al.}(2023)\citenamefont {Beeman}
  \emph {et~al.}}]{Beeman:2022wun}%
  \BibitemOpen
  \bibfield  {author} {\bibinfo {author} {\bibfnamefont {J.~W.}\ \bibnamefont
  {Beeman}} \emph {et~al.},\ }\href {\doibase 10.1016/j.apradiso.2023.110704}
  {\bibfield  {journal} {\bibinfo  {journal} {Appl. Radiat. Isot.}\ }\textbf
  {\bibinfo {volume} {194}},\ \bibinfo {pages} {110704} (\bibinfo {year}
  {2023})},\ \Eprint {http://arxiv.org/abs/2206.05116} {arXiv:2206.05116
  [physics.ins-det]} \BibitemShut {NoStop}%
\end{thebibliography}%

\end{document}